\documentclass[useAMS,usenatbib]{mn2e}
\usepackage{epsfig}

\def\bo{{\rm B}}
\def\sb{{\rm SB}}
\def\ph{{\rm ph}}
\def\th{{\rm Th}}
\def\rr{{\rm r}}
\def\eff{{\rm eff}}
\def\acc{{\rm acc}}
\def\ob{{\rm obs}}
\def\emm{{\rm em}}
\def\k{{\rm K}}
\def\arctanh{{\rm arctanh}}
\def\arccosh{{\rm arccosh}}

\def\com{{\rm com}}
\def\p{{\rm peak}}

\def\tot{{\rm total}}
\def\lo{{\rm local}}
\def\m{{\rm m}}
\def\c{{\rm c}}
\def\tr{{\rm tr}}

\def\sw{{\em Swift}}
\def\integ{{\em INTEGRAL}}
\def\cgro{{\em CGRO}}
\def\kw{Konus/{\em WIND}}

\title[Gamma-Ray Burst Precursors] 
{Gamma-Ray Burst Precursors as the Remnant of the Thermal Radiation 
Initially Trapped in the Fireball}
\author[Li-Xin Li]{Li-Xin Li\thanks{E-mail: lxl@mpa-garching.mpg.de}\\
Max-Planck-Institut f\"ur Astrophysik, 85741 Garching, Germany}

\begin{document}


\date{Accepted 2007 June 6. Received 2007 May 31; in original form 2007 
March 7}

\pagerange{\pageref{firstpage}--\pageref{lastpage}} \pubyear{2006}

\maketitle

\label{firstpage}

\begin{abstract}
In the standard fireball model of gamma-ray bursts (GRBs), the fireball 
starts with an optically thick phase. As it expands, the fireball becomes 
optically thin at some stage. The thermal radiation trapped in the originally 
opaque fireball then leaks out, producing a transient event. The appearance
of the event is investigated in the framework of a homogeneous, spherically 
symmetric, and freely expanding fireball produced instantly by an explosive 
process without continuous injection of mass and energy. We find that, 
generally, the event has a time-duration shorter than that of the main burst, 
which is presumably produced by the internal shock after the fireball becomes 
optically thin. The event is separated from the main burst by a quiescent 
time-interval, and is weaker than the main burst at least in a high energy 
band. Hence, the event corresponds to a GRB precursor. The precursor event 
predicted by our model has a smooth and FRED (Fast Rise and Exponential Decay)
shape lightcurve, and a quasi-thermal spectrum. Typically, the characteristic 
blackbody photon energy is in the X-ray band. However, if the distortion of 
the blackbody spectrum by electron scattering is considered, the 
characteristic photon energy could be boosted to the gamma-ray band. Our 
model may explain a class of observed GRB precursors---those having smooth 
and FRED-shape lightcurves and quasi-thermal spectra.
\end{abstract}


\section{Introduction}
\label{intro}

A standard model for gamma-ray bursts (GRBs) has been the fireball model 
\citep{goo86,pac86}. In this scenario, it is assumed that by whatever a
process, a radiation-dominated, optically thick, and baryon-poor plasma 
fluid is suddenly produced in a compact volume. The radiation field drives 
the fireball into relativistic expansion, so that a significant fraction of 
the initial energy of the radiation is converted to the kinetic energy of the 
fireball \citep{pac90,she90}. At the end of acceleration, the fireball starts
to expand freely with a constant Lorentz factor $\ga 100$ (Kobayashi, Piran 
\& Sari 1999). Later on, the fireball becomes optically thin and the thermal
radiation trapped originally in the fireball starts to leak out. As the 
fireball becomes sufficiently large, its kinetic energy is converted to the 
prompt gamma-ray emission through the internal shock and the afterglow 
emission through the external shock \citep{ree92,ree94,pac94}. For a review 
on the fireball model and the internal/external shocks, see \citet{mes06}, 
\citet{pir99,pir04}, and \citet{zha04}.

In many ways a GRB fireball is like a cosmological Big Bang \citep{pee93}. 
Both theories assume that the event (the GRB and the Universe) starts with 
a state of a radiation-dominated plasma of very high temperature, in which 
the production of electron-positron pairs is important, and the plasma is 
optically thick. The energy of the radiation drives the expansion of the 
plasma. The plasma and the radiation cool down as the fireball expands. At 
some moment, the mass density of the plasma becomes low enough so that the 
plasma becomes optically thin to the photons trapped in it, and the fireball 
(of the GRB or of the Universe) undergoes a transition from an opaque phase 
to a transparent phase. 

However, a GRB fireball also differs from the Big Bang. The Universe is 
homogeneous and isotropic at a very high degree. At the time of 
recombination the fluctuation in the temperature of the cosmic microwave 
background (CMB) radiation relative to the mean temperature is only 
$\sim 10^{-5}$ \citep{ben96,smo92}. In contrast, the GRB fireball could be 
highly inhomogeneous and anisotropic. Indeed, the fluctuation in the Lorentz 
factor of the fireball is required for the internal shock model to work 
\citep{ree94,pac94}. The Universe is a closed system so that the total energy 
in it must be conserved. However, the GRB fireball interacts with the
surrounding matter and photons are radiated away from its surface. The central
engine of the GRB may also continue pumping energy into the fireball even 
after the prompt GRB emission has ended \citep[and references therein]{bur07}. 
In addition, it is usually assumed that the Universe is infinite, but of
course the GRB fireball has a finite volume.

A remarkable success of the Big Bang theory has been the prediction of the 
existence of a blackbody CMB of temperature $\approx 3$~K in today's Universe
\citep{gam48a,gam48b,alp48} and its detection \citep{pen68,smo92}. The CMB
is in fact the cooled remnant of the primeval fireball---an echo of the Big 
Bang. Because of the expansion of the Universe, the temperature of the CMB is
redshifted and is hence very low as the CMB photons reach an observer of
today. Similarly, we expect that the GRB fireball has also a remnant of the
radiation initially trapped in the fireball, and that the remnant would have
a quasi-thermal spectrum. However, because of the fact that the surface of 
the fireball is assumed to be moving relativistically towards an observer who 
detects the GRB, the temperature of the radiation as measured by the observer 
would be significantly boosted by the Doppler effect. 

Because of the fact that the radiation in the fireball has a finite energy, 
the remnant event must have a finite duration. The event starts when the 
fireball is still optically thick, while the main burst takes place when the 
fireball is already optically thin. Hence, the remnant event must occur 
before the main burst, with a smaller distance from the GRB central engine 
than the main burst. As we will see, the remnant event often has a shorter 
duration than the main burst (due to the remnant event's smaller distance 
from the central engine), and is separated from the main burst by a quiescent 
period of time. Since it is produced by the emission from the photosphere of 
the fireball, the remnant event should have a spectrum that is dominated by 
a quasi-thermal component. Hence, at least in a high energy band, the remnant 
event should look weaker than the main burst. Therefore, the remnant event 
should be observed as a {\em precursor} of the GRB. 

The aim of the paper is to quantitatively investigate the properties of the 
GRB precursors as the remnant of the radiation initially trapped in the 
fireball and look for their possible observational consequences.

In at least several cases, precursors of GRBs have been unambiguously 
detected. GRB 030406, a burst that was detected out side of the field of view 
of the International Gamma-Ray Astrophysics Laboratory (\integ), had a 
precursor that occurred $\sim 50$~s before the main burst \citep{mar06}. 
GRB 041219a, a burst that was detected by the \integ\, Burst Alert System 
(IBAS), had a precursor that occurred $\sim 260$~s before the main burst 
\citep{ves05,mcb06}. GRB 050820a triggered the Burst Alert Telescope (BAT) 
onboard \sw\, by a precursor, and \kw\, by the main burst. The precursor 
occurred $\sim 200$~s before the main burst, and the entire duration of the
burst is $\sim 600$~s \citep{cen06}. GRB 060124, also detected by both 
BAT/\sw\, and \kw, had a precursor that occurred $\sim 500$~s before the 
main burst \citep{rom06}. The entire duration is $\sim 800$~s, making 
GRB 060124 one of the longest bursts. Another interesting case is 
GRB 061121, detected by BAT/\sw\, (also by \kw\, and {\em RHESSI}), which had 
a precursor that occurred $\sim 60$~s before the main burst 
\citep{bel06,fen06,gol06,pag06,pag07}.

These observed gamma-ray precursors have the following characters: (1) The 
precursor is separated from the main burst by a long period of quiescent time; 
the time separation is comparable to the duration of the main burst. (2) The 
precursor is weaker than the main burst, and has a shorter time-duration. 
(3) The precursor often has a smooth and FRED (Fast Rise and Exponential 
Decay) shape lightcurve. (GRB 050820a is an exception, whose precursor has 
a lightcurve with at least two peaks.) (4) The precursor has a much softer 
spectrum than the main burst, and in at least two cases (GRB 030406 and 
GRB 041219a) the spectrum can be fitted by a blackbody or a blackbody plus 
a power-law.

On the other hand, \citet{kos95} have found that about 3 percent of the GRBs 
detected by the Burst and Transient Spectrometer Experiment (BATSE) on the 
Compton Gamma-Ray Observatory (\cgro) have precursors. With a precursor
definition different from that of \citet{kos95}, \citet{laz05} has found that 
about 20 percent of the long-duration BATSE GRBs have evidence of precursor 
emission.\footnote{\citet{laz05} adopted a definition for a GRB precursor
that is in favor of weak precursor emissions and put no limit on the
time separation between the precursor and the main burst, hence he found
more precursors than \citet{kos95}.}
Soft precursor activities in X-rays have been detected in 
a number of GRBs by {\em Ginga} \citep{mur91,mur92} and WATCH onboard 
{\em GRANAT} \citep{saz98}. The precursors detected by {\em Ginga} have a 
thermal spectrum with a temperature $\sim 1$--$2$ keV. 

On the thermal emission in GRBs, we would also mention that 
\citet{ryd04,ryd05} has found that up to 30 percent of long-duration GRBs 
detected by BATSE/\cgro\, have a spectrum that can be interpreted as 
combination of a thermal peak plus a power-law component. But his results 
refer to the prompt gamma-ray emission in the main burst, not the precursor 
emission.

In this paper, we consider a very simple model for the GRB fireball. By 
assumption, the fireball is homogeneous and spherically symmetric, and expands 
with a constant and relativistic speed. The fireball contains a thermal 
radiation field, but the energy density of the radiation is not large enough 
to affect the dynamics of the fireball. The rest mass and the kinetic energy 
of the fireball are conserved. For the radiation field in the fireball, the
change in the number of photons is caused only by the emission from the 
photosphere. Hence, the number of photons is an adiabatically conserved
quantity. The last condition requires that the electron-positron pairs
have already annihilated. Thus, in addition, we assume that the opacity
in the fireball is given by the constant Thompson opacity.

Our model corresponds to the post-acceleration phase of the GRB outflow. This
assumption is justified by the fact that the radius of the fireball at the
time of photosphere emission is much larger than the radius at the end of 
acceleration \citep[and Section \ref{char} in the present paper]{dai02}. Our 
model is similar to that of \citet{goo86} in several aspects: both models 
assume a freely expanding and optically thick fireball in which an amount of 
energy was generated instantly at a beginning time, and the observable event 
is determined by the evolution of the fireball and the radiation in it. There 
is no continuous injection of energy or mass, unlike the steady wind model of 
\citet{pac86,pac90}. However, our model also differs from that of 
\citet{goo86}. In the model of \citet{goo86}, the fireball is very hot and
the process of pair production is very important, and the photosphere was not
explicitly included in the calculation. While in our model, the fireball is 
relatively cool and hence the process of pair production can be ignored 
(see Section \ref{char} for a justification of this assumption), and the 
photosphere is calculated in details. In addition, the model of \citet{goo86}
[as well as that of \citet{pac86,pac90}] was aimed to interpret the prompt
emission of GRBs (the main bursts), while our model is aimed to interpret 
the precursor emission of GRBs. This leads to a cooler fireball since a GRB
precursor usually has a spectrum that is softer than the main burst.

The paper is organized as follows. In Section \ref{milne}, we describe the
geometry and the kinematics of a GRB fireball, the kinematics of a radiation 
field in it, and the structure of the photosphere. In Section \ref{precursor}, 
we describe a formalism for calculating the properties of the precursor event 
of a GRB arising from the emission by the photosphere, including the 
luminosity, the blackbody spectrum, and the photon rate observed by a remote 
observer. In Section \ref{numer}, we present our numerical results. In 
Section \ref{char}, we derive some scaling relations for the characteristic 
quantities of the precursor, including the characteristic time scale, the 
characteristic total energy, and the characteristic photon energy. We also 
give brief justification for some key assumptions in our model. In Section 
\ref{discussion}, we discuss the effect of jet collimation, the dependence of 
our results on the energy band of the detector, and the effect of spectrum
distortion by electron scattering. In Section \ref{concl}, we summarize the 
results and draw our conclusions.

In Appendix \ref{const_vel}, we present the simplified results for a limiting 
case: a photosphere with a constant expansion velocity, which applies to the 
beginning part of the precursor.

We remark that thermal precursors of GRBs in the internal shock model have 
been previously studied by \citet{dai02}. The thermal emission from a GRB 
photosphere and its effect on the prompt spectrum of a GRB have 
been investigated by \citet{mes00}; \citet{mes02}; \citet{ram05}; 
\citet{ree05}; Pe'er, M\'esz\'aros \& Rees (2006); \citet{pee07}; and 
Thompson, M\'esz\'aros \& Rees (2007). The
emission from a photosphere of a Poynting flux dominated magnetized fireball 
and its relation to the GRB precursor or the prompt emission of the main 
burst has been discussed by \citet{lyu00}; \citet{gia06}; and \citet{gia07}. 
Our model differs from that adopted in their papers. In their work, they have 
assumed a steady wind model for the fireball, in which mass and energy are 
continuously injected into the wind at the center. And, the expansion of the
fireball in our model was driven by radiation, not by the Poynting flux as
in the magnetic models.

Bianco, Ruffini \& Xue (2001) have calculated the thermal emission from a
fireball dominated by electron-positron pairs arising from the quantum vacuum 
polarization process around a charged black hole. The results have been
applied to interpretation of the precursor emission, the main burst, as well 
as the afterglow of GRBs in a unified way \citep{ruf01,ruf02,ruf05}.

We also remark that thermal precursors produced by the shock wave (or jet) 
breakout of the progenitor star in the collapsar model of GRBs have been 
studied by MacFadyen \& Woosley (1999); MacFadyen, Woosley \& Heger (2001); 
Ramirez-Ruiz, MacFadyen \& Lazzati (2002); and Waxman \& M\'esz\'aros (2003).

Finally, whenever an observer is referred to in the paper, we ignore the
cosmological effect. That is, we assume that the Universe is Euclidean. The 
correction of the cosmological effect is straightforward, which mainly
includes three factors: cosmological redshift to the observed photon
energy, time dilation in the observed lightcurve duration, and that the 
GRB distance appearing in our formulae should be interpreted as the 
appropriate cosmological distance.

\section{The Fireball as a Milne Universe}
\label{milne}

A freely expanding, homogeneous, and spherically symmetric fireball is like a 
Milne universe (see, e.g., Rindler 1977), except that a fireball has a finite 
volume but the Milne universe has an infinite volume. 

Let us denote the time in the GRB's rest frame by $t$, and the spherical 
coordinates in it by $\{r,\theta,\phi\}$. The center of the fireball is at
$r=0$. Then, the Minkowski metric is
\begin{eqnarray}
	ds^2 = - c^2 dt^2 + dr^2 + r^2 d\Omega^2 \;,  \label{ds1}
\end{eqnarray}
where $d\Omega^2 \equiv d\theta^2+\sin^2\theta d\phi^2$ is the metric on a 
two-dimensional sphere of a unit radius.

\begin{figure}
\vspace{2pt}
\includegraphics[angle=0,scale=0.69]{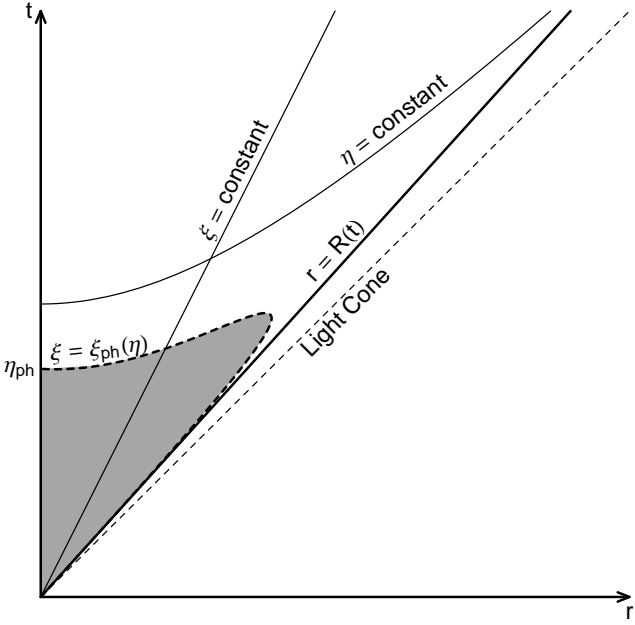}
\caption{The spacetime diagram of a fireball as a Milne universe. Each
point represents a two-sphere. The hypersurface $\xi=\xi_\ph(\eta)$ defines
the photosphere (Sec.~\ref{photo}), within which (the shaded region) the
photon optical depth is larger than unity. The photosphere hypersurface 
becomes spacelike when $\eta>\eta_0/\sqrt{2}$, and ends at $\eta=\eta_\ph$.
}
\label{fireball}
\end{figure}

Define the coordinates $\eta$ and $\xi$ by
\begin{eqnarray}
	t=\eta \cosh\xi \;, \hspace{1cm} r= c\eta \sinh\xi \;,
	\label{tr}
\end{eqnarray}
the metric can then be rewritten as the Milne metric
\begin{eqnarray}
	ds^2 = - c^2 d\eta^2 + c^2 \eta^2 \left(d\xi^2 + \sinh^2\xi\,
		d\Omega^2\right) \;.
	\label{ds2}
\end{eqnarray}

The trajectory of a particle with a constant radial velocity $v=\beta c$ is a 
straight line defined by
\begin{eqnarray}
	r = c\beta t \;,  \label{r_t}
\end{eqnarray}
aside from that $\theta=\mbox{constant}$ and $\phi=\mbox{constant}$. 
Equation (\ref{r_t}) demonstrates that, at any moment of constant $t$, we have
$v\propto r$. This profile of expansion velocity is typical for explosive 
events, e.g. supernovae. 

By equation (\ref{tr}) we have that, along the trajectory of the particle,
\begin{eqnarray}
	\xi = \arctanh \beta \;, \hspace{1cm} \eta = \gamma^{-1} t \;, 
	\label{t_eta}
\end{eqnarray} 
where $\gamma\equiv \left(1-\beta^2\right)^{-1/2}= \cosh\xi$ is the Lorentz 
factor of the particle.

Therefore, $\eta$ represents the proper time of the particle, and $\xi$ 
measures the spatial velocity of the particle. The spatial distance in the 
radial direction on a hypersurface $\Sigma_\eta$ defined by a constant $\eta$ 
is, by equation~(\ref{ds2}), $l = c \eta \int_0^\xi d\xi = c \eta\xi$\,.

In Fig.~\ref{fireball}, we show the spacetime diagram of the fireball. The
fireball has an outer boundary at $r=R(t) = c\beta_R t$, where the expansion
velocity is $v_R=c\beta_R$, and the Lorentz factor is $\Gamma = \left(1-
\beta_R^2\right)^{-1/2}$.

\subsection{The Kinematics of the Fireball}
\label{kine}

The comoving spatial volume of a sphere of a coordinate radius $\xi$ on the 
hypersurface $\Sigma_\eta$ is
\begin{eqnarray}
	V_\com(\xi) = 4\pi c^3\eta^3 \int_0^\xi \sinh^2\xi\, d\xi 
		= \pi c^3\eta^3 (\sinh 2\xi - 2\xi) \;.
	\label{vol_com}
\end{eqnarray}
For a constant $\xi$, the comoving volume is $\propto \eta^{3}$.

In the non-relativistic limit $\beta^2 \approx \xi^2 \ll 1$, we have $V_\com 
\approx 4\pi r^3/3$. In the ultra-relativistic limit
$\gamma=\cosh\xi\gg 1$, we have 
\begin{eqnarray}
	V_\com \approx 2\pi\gamma^2 c^3\eta^3 \approx
		\frac{2\pi}{\gamma} r^3 \;. 
	\label{vol_com_rel}
\end{eqnarray}

The fireball has a uniform comoving mass density $\rho = \rho(\eta) \propto 
\eta^{-3}$. The total rest mass contained in the fireball is then
\begin{eqnarray}
	M = \rho V_\com\left(\xi_R\right) = \frac{M_0}{4} \left(\sinh 2\xi_R 
		- 2\xi_R\right) \;,
	\label{mass}
\end{eqnarray}
where $\xi_R \equiv \arccosh\Gamma$ is the value of $\xi$ at the outer
boundary of the fireball, and the constant reference mass
\begin{eqnarray}
	M_0 \equiv 4\pi\rho\eta^3c^3  \;. \label{M0}
\end{eqnarray}

The kinetic energy of the mass contained in a comoving volume element 
$dV_\com$ bounded by two neighbored spheres of constant radii $\xi$ and 
$\xi+d\xi$ is
\begin{eqnarray}
	d E_\k = (\gamma-1) \rho c^2 dV_\com
		= (\cosh\xi-1) M_0 c^2 \sinh^2\xi d\xi \;,
\end{eqnarray}
which is independent of $\eta$. Hence, the total kinetic energy of the fireball
is
\begin{eqnarray}
	E_\k &=& \int_{\xi=0}^{\xi_R} d E_\k \nonumber\\
		&=& M_0 c^2 \left(\frac{1}{3} \sinh^3\xi_R -
		\frac{1}{4} \sinh 2\xi_R + \frac{1}{2}\xi_R\right) \;.
	\label{kinetic}
\end{eqnarray}

Equation~(\ref{kinetic}) can also be derived as follows. The stress-energy
tensor of the particle fluid is $T_{ab} = \rho c^2 d\eta_a d\eta_b$. The 
energy density measured by a rest observer is $T_{ab}\left(\partial/\partial 
t\right)^a \left(\partial/\partial t\right)^b = \gamma^2\rho c^2$. The total 
energy contained in a hypersurface $\Sigma_t$ defined by $t=\mbox{constant}$ 
(i.e., a three-space in the rest frame) is thus
\begin{eqnarray}
	E = 4\pi \int_0^R \gamma^2\rho c^2 r^2 dr = \frac{M_0}{c} 
		\int_0^R \gamma^2 \eta^{-3} r^2 dr \;,
	\label{et}
\end{eqnarray}
where the integral is evaluated on $\Sigma_t$. Substituting $r=c\beta t$ and 
$\eta= \gamma^{-1} t$ into equation~(\ref{et}) and treating $t$ as a constant,
we get
\begin{eqnarray}
	E = M_0 c^2 \int_0^{\beta_R} \gamma^5 \beta^2 d\beta
		= \frac{1}{3} M_0 c^2 \sinh^3 \xi_R \;,
	\label{et2}
\end{eqnarray}
where $\beta =\tanh\xi$ and $\gamma = \cosh\xi$ have been used. From 
equations~(\ref{mass}) and (\ref{et2}), $E-Mc^2$ is just the kinetic energy 
$E_\k$ in equation~(\ref{kinetic}).

When $\beta_R^2\ll 1$ ($\Gamma\approx 1$, the non-relativistic limit), we have 
$M\approx M_0\beta_R^3/3 \approx 4\pi \rho R^3/3$ and $E_\k \approx M_0 c^2 
\beta_R^5/10 \approx 3 M v_R^2/10$. As expected, all these results return to 
the Newtonian values.

When $\Gamma=\cosh\xi_R \gg 1$ ($\beta_R\approx 1$, the ultra-relativistic
limit), we have
\begin{eqnarray}
	M \approx \frac{1}{2} M_0 \Gamma^2 \;,  \label{mass_rel}
\end{eqnarray}
and
\begin{eqnarray}
	E\approx E_\k \approx \frac{2}{3}\Gamma Mc^2 \;. \label{ener_rel}
\end{eqnarray}

\begin{figure}
\vspace{2pt}
\includegraphics[angle=0,scale=0.465]{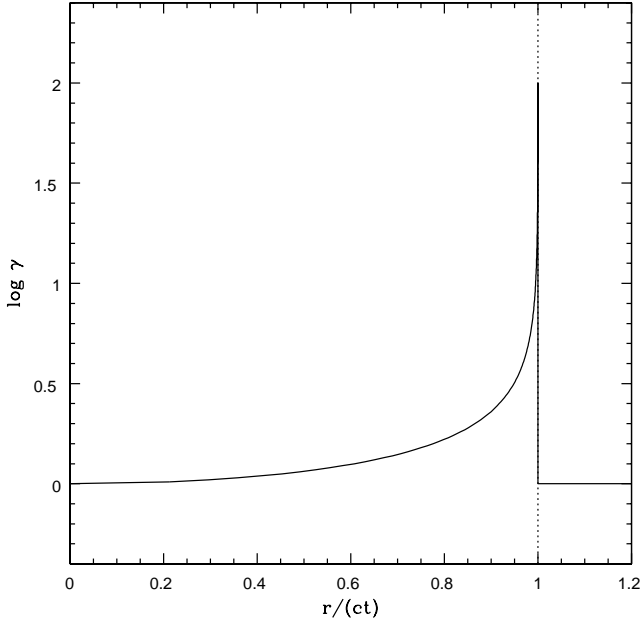}
\caption{The Lorentz factor of particles in the fireball as a function of
radius, at any moment of constant $t$: $\gamma = \left[1-r^2/(ct)^2
\right]^{-1/2}$. At the outer boundary of the fireball (the dotted line), the 
Lorentz factor $\gamma = \Gamma = 100$.
}
\label{gamma_r}
\end{figure}

In the comoving frame the fireball has a uniform density. However, in the
rest frame of the GRB, the density of the fireball increases with radius
because of the Lorentz contraction. This effect is particularly important near 
the outer boundary of the fireball, where the Lorentz factor is 
dramatically large in the ultra-relativistic case (Fig. \ref{gamma_r}). 

In the rest frame, the number density of the fireball particles is $\propto
\gamma\rho\propto \gamma^4$ on $\Sigma_t$. The total energy density is 
$\gamma^2\rho c^2 \propto\gamma^5$. Hence, in the rest frame both mass and 
energy are concentrated to a thin spherical shell at the outer boundary. 
If we consider a thin spherical shell defined by $R-\Delta R < r <R$ with a 
thickness (in the rest frame) 
\begin{eqnarray}
	\Delta R = \frac{R}{2\Gamma^2} \ll R \;,  \label{delta_r}
\end{eqnarray}
then by equations (\ref{mass_rel}) and (\ref{ener_rel}), half of the total 
mass and $64.6$ percent of the total kinetic energy are contained in the 
shell.

\subsection{The Radiation Field in the Fireball}
\label{rad}

Assume that the fireball contains a uniform radiation field with a comoving
energy density $e_\rr$ and a pressure $p_\rr = e_\rr/3$. From the conservation 
of energy, $e_\rr\propto \eta^{-4}$. The total energy of the radiation is 
not larger than the kinetic energy of the fireball, so that the assumption 
of a freely expanding fireball is valid ($\lambda<1$, eqs. \ref{model_cond} 
and \ref{lam2}).

\begin{figure}
\vspace{2pt}
\includegraphics[angle=0,scale=0.7]{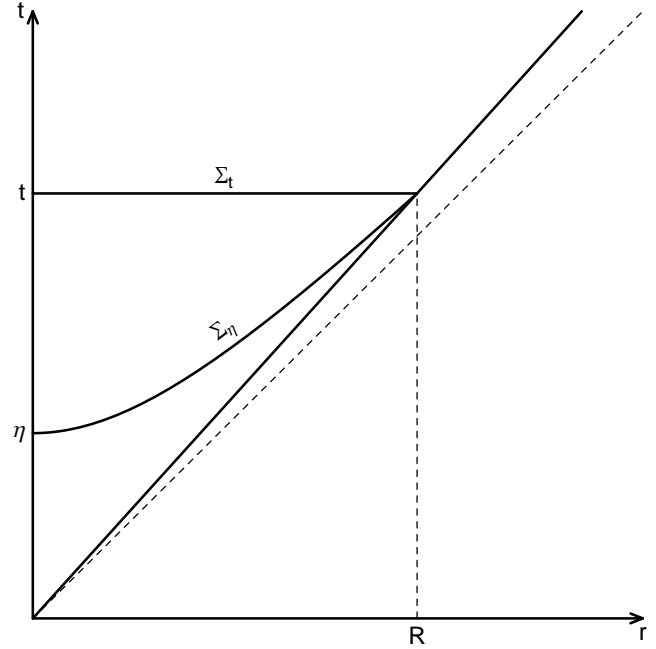}
\caption{The hypersurface on which the total energy of radiation is defined.
The energy in the comoving frame, $E_{\rr,\com}$, is defined on $\Sigma_\eta$
($\eta=\mbox{constant}$). The energy in the observer's frame, $E_\rr$, is 
defined on $\Sigma_t$ ($t=\mbox{constant}$). $\Sigma_t$ and $\Sigma_\eta$
intersect at $r=R$ (the outer boundary of the fireball), so that $E_\rr$
and $E_{\rr,\com}$ are related by equation (\ref{gr_eff}). (The diagonal
dashed line represents the light cone.)
}
\label{rad_ener}
\end{figure}

The total energy of the radiation defined in the comoving frame is
\begin{eqnarray}
	E_{\rr,\com} = e_\rr V_\com = \pi e_\rr c^3 
		\eta^3\left(\sinh 2\xi_R-2\xi_R\right) \;.
	\label{er_com}
\end{eqnarray}
Since $e_\rr\propto \eta^{-4}$, we have $E_{\rr,\com}\propto \eta^{-1}$. As 
the fireball expands, the total energy of the radiation decreases, caused by 
the fact that the pressure of the radiation does work. However, since the 
number density of photons $n_\rr\propto \eta^{-3}$, the total number of 
photons in the fireball is conserved.\footnote{Strictly speaking, the number 
of photons is an adiabatically conserved quantity since the emission of
photons by the photosphere reduces the number of photons in the fireball
gradually.}

The stress-energy tensor of the radiation field is $T_{ab} = \left(e_\rr+ 
p_\rr\right) d\eta_a d\eta_b + \left(p_\rr/c^2\right) g_{ab}$, where $g_{ab}$ 
is the spacetime metric. The energy density in the rest frame is then $T_{ab}
\left(\partial/\partial t\right)^a \left(\partial/\partial t\right)^b = 
\gamma^2 e_\rr+\left(\gamma^2 -1\right) p_\rr = \left(4\gamma^2-1\right)
e_\rr/3$. The total energy of the radiation measured in the rest frame 
is then
\begin{eqnarray}
	E_{\rr} = \frac{1}{3}\int_{\Sigma_t} \left(4\gamma^2-1\right)
		e_\rr 4\pi r^2 dr \;.
\end{eqnarray}

Using $e_\rr\propto \eta^{-4}$, $\eta=\gamma^{-1} t$, and $r=c\beta t$, we get
\begin{eqnarray}
	E_{\rr} = \frac{4\pi}{3} \left(e_\rr \eta^4\right) c^3
		t^{-1} \int_{\Sigma_t} \left(4\gamma^2-1\right)\gamma^4 
		\beta^2 d\beta \;.
\end{eqnarray}
Since $\beta=\tanh\xi$ and $\gamma=\cosh\xi$, we can work out the integral
and obtain
\begin{eqnarray}
	E_{\rr} = \frac{4\pi}{3} \left(e_\rr \eta^4\right) c^3
		t^{-1}\cosh\xi_R\sinh^3\xi_R \;.
\end{eqnarray}
Hence, $E_\rr\propto t^{-1}$.

If we define $\Sigma_\eta$ with the requirement that $\Sigma_\eta$ intersects
$\Sigma_t$ at $r=R$ (i.e., $\eta = t/\Gamma = t/\cosh\xi_R$; see 
Fig.~\ref{rad_ener}), we can compare the $E_\rr$ defined on $\Sigma_t$ and the
$E_{\rr,\com}$ defined on $\Sigma_\eta$. We have then the effective Lorentz 
factor of the radiation field
\begin{eqnarray}
	\Gamma_{\rr,\eff} &\equiv& \frac{E_\rr}{E_{\rr,\com}}
		= \frac{4}{3}\frac{\sinh^3\xi_R}{\sinh 2\xi_R -2\xi_R} \;.
	\label{gr_eff}
\end{eqnarray}

When $\Gamma\approx 1$, we have $\Gamma_{\rr,\eff}\approx 1$, and $E_\rr
\approx E_{\rr,\com} \approx (4\pi/3) e_\rr R^3$. When $\Gamma \gg 1$, we 
have $\Gamma_{\rr,\eff} \approx 2\Gamma/3$ and 
\begin{eqnarray}
	\left.E_\rr\right|_{\Sigma_t} \approx \frac{2}{3} \Gamma 
		\left. E_{\rr,\com}\right|_{\Sigma_\eta} \approx
		\frac{4\pi}{3} e_\rr R^3 \;, 
	\label{e_rr_rel}
\end{eqnarray}
where $e_\rr = e_\rr(\eta)$ is evaluated on $\Sigma_\eta$, and $R \approx
c t$ is the radius of the fireball at time $t$ (Fig.~\ref{rad_ener}).

At any moment of constant $t$, we have $\left.E_\rr\right|_{\Sigma_t} \propto
e_\rr \propto \eta^{-4} \propto \gamma^4$. Hence, $75$ percent of the total
radiation energy (defined in the rest frame of the GRB) is contained in the
thin shell at the outer boundary of the fireball with a thickness given by
equation (\ref{delta_r}).

We can also compare the total energy of radiation on $\Sigma_t$ to the total 
kinetic energy of particles, in the ultra-relativistic limit. By 
equations (\ref{vol_com_rel}), (\ref{mass}), and (\ref{ener_rel}), we have
\begin{eqnarray}
	E_\k \approx \frac{4\pi}{3}\rho R^3 c^2 \;, \label{ek_rho}
\end{eqnarray}
where $\rho$ is defined on the $\Sigma_\eta$ in Fig.~\ref{rad_ener}. Hence, 
by equation (\ref{e_rr_rel}), we have
\begin{eqnarray}
	\lambda\equiv\left.\frac{E_\rr}{E_\k}\right|_{\Sigma_t} 
		= \left.\frac{e_\rr}{\rho c^2}\right|_{\Sigma_\eta} \;.
	\label{model_cond}
\end{eqnarray}

To make the model self-consistent, we must require that $\lambda< 1$. In our 
model, $\rho\propto\eta^{-3}$ but $e_\rr \propto \eta^{-4}$, and so $\lambda
\propto \eta^{-1}\propto t^{-1}$. Hence, if the condition $\lambda< 1$ is 
satisfied at some moment, it will remain being satisfied afterwards. As 
$\eta\rightarrow 0$ we have $e_\rr/\rho c^2 \rightarrow \infty$ and $\lambda
\rightarrow\infty$. Hence, there must exist a transition time $\eta_\acc$ 
defined by $\lambda(\eta_\acc)=1$. When $\eta<\eta_\acc$, the fireball is 
accelerated by the radiation field. Our model only applies to the 
free-expansion phase of $\eta>\eta_\acc$.

\subsection{The Photosphere of the Fireball}
\label{photo}

To calculate the photon optical depth in the fireball, we need to consider 
the geodesics of a photon. Assume that a photon is emitted from a point on 
a sphere of a comoving coordinate radius $\xi = \xi_1$ at comoving time 
$\eta_1$, moving outwards in the radial direction. The geodesics of the
photon is described by $r = r_1 + c(t-t_1)$ in terms of $\{t,r\}$, where 
$t_1$ and $r_1$ are related to $\eta_1$ and $\xi_1$ by equation~(\ref{tr}).
In terms of $\{\eta,\xi\}$, the geodesics can be written as
\begin{eqnarray}
	\eta = \eta_1 e^{\xi-\xi_1} \;. \label{geod}
\end{eqnarray}

Assume that the opacity in the fireball, $\kappa$, is a constant. Then, the
optical depth along the trajectory of the photon is \citep{nov73}
\begin{eqnarray}
	\tau = \kappa \int \rho dl = \frac{\kappa M_0}{4\pi c^2} 
		\int_{\xi_1}^{\xi_R} \eta^{-2} d\xi \;,   
	\label{tau_0}
\end{eqnarray}
where $dl = c\eta d\xi$ is radial distance element in the comoving frame, and
the integral is along the geodesics of the photon. 

Submitting equation (\ref{geod}) into equation (\ref{tau_0}), we get
\begin{eqnarray}
	\tau = \frac{\kappa M_0}{8\pi c^2\eta_1^2}\left[1-
		e^{-2\left(\xi_R-\xi_1\right)}\right]  \;.
	\label{tau_1}
\end{eqnarray}

When $\xi_1= 0$ and $\beta_R\ll 1$, we get $\tau\approx 3\kappa M/4\pi R^2$, 
returning to the Newtonian result.

When $\xi_1= 0$ and $\Gamma\gg 1$, we get 
\begin{eqnarray}
	\tau\approx \frac{\kappa M_0}{8\pi c^2\eta_1^2} \approx
		\frac{\kappa M}{4\pi R_1^2\Gamma^2} \;,
	\label{tau2}
\end{eqnarray}
where $R_1\equiv c\eta_1$ is the radius of the fireball at the time when the 
photon is emitted.

Since $\tau\propto\eta_1^{-2}$, at very early time we must have $\tau\gg 1$
and the fireball must be optically thick. We define the photosphere of the
fireball as a hypersurface determined by $\tau = 1$. Then, by equation
(\ref{tau_1}), we get the equation for the photosphere
\begin{eqnarray}
	\xi_\ph = \xi_R +\frac{1}{2}\ln\left(1-\frac{\eta^2}{\eta_0^2}
		\right) \;,
	\label{rph_eq}
\end{eqnarray}
where
\begin{eqnarray}
	\eta_0 \equiv \left(\frac{\kappa M_0}{8\pi c^2}\right)^{1/2} 
		= \left(\frac{3\kappa E_\k}{8\pi c^4\Gamma^3}
		\right)^{1/2} \;. 
	\label{eta0}
\end{eqnarray}

The photosphere hypersurface is shown in Fig.~\ref{fireball}. It starts from 
$\eta=0$ as a timelike hypersurface, becomes spacelike when $\eta>\eta_0/
\sqrt{2}$ and ends at $\eta=\eta_\ph$, where
\begin{eqnarray}
	\eta_\ph \equiv \eta_0\left(1-e^{-2\xi_R}\right)^{1/2} \;.
	\label{eta_ph}
\end{eqnarray}
By the time $\eta = \eta_\ph$, the radius of the photosphere shrinks to zero. 
At $\eta\ll \eta_0$, we have $\xi_\ph\approx \xi_R$.

Submitting equation (\ref{eta0}) into equation (\ref{tau_1}), we get
\begin{eqnarray}
	\tau = \frac{\eta_0^2}{\eta_1^2}\left[1-e^{-2\left(\xi_R -
		\xi_1\right)}\right] \;. 
	\label{tau_eta}
\end{eqnarray}

In the ultra-relativistic limit $\Gamma = \cosh\xi_R\gg 1$, we have that
$\eta_\ph\approx \eta_0\left(1-1/8\Gamma^2\right)\approx \eta_0$, and $\tau
\approx \eta_0^2/\eta_1^2$ when $\xi_1=0$.

Although the mass and the energy are concentrated in a thin shell at the
outer boundary of the fireball (Sections \ref{kine} and \ref{rad}), for a
photon emitted from the center of the fireball the dominant contribution to
the total optical depth comes from the central region. By equation 
(\ref{tau_1}) (with $\xi_1=0$ and $e^{\xi_R}\gg 1$), the optical depth is 
$\tau/2$ at $\xi = \ln 2^{1/2} \approx 0.3466$, where $\gamma = 3\sqrt{2}/
4\approx 1.0607$. This is caused by the fact that the mass density decreases 
quickly as the fireball expands and the optical depth is Lorentz invariant.

\begin{figure}
\vspace{2pt}
\includegraphics[angle=0,scale=0.463]{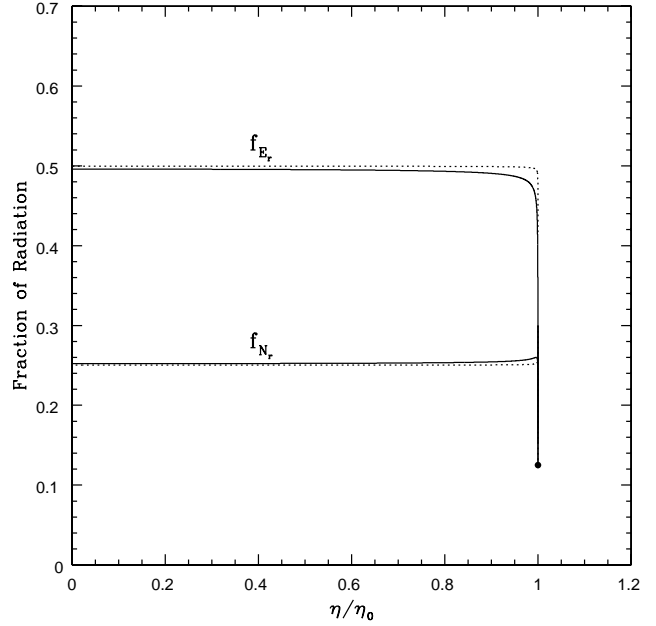}
\caption{The fractions $f_{E_\rr}$ (eq. \ref{fr1}, two upper lines) and 
$f_{N_\rr}$ (eq. \ref{fr2}, two lower lines) as functions of $\eta$---the 
comoving time on the photosphere. The solid curves are for $\Gamma=100$. 
The dotted curves are for $\Gamma = 1000$. The maximum of $\eta$ is 
$\eta_\ph\approx \eta_0$, defined by equation (\ref{eta_ph}) and marked by 
the dot. At $\eta=\eta_\ph$, we have $f_{E_\rr} = f_{N_\rr} = 1/8$. 
}
\label{energy_tau}
\end{figure}

For a photon to arrive at the photosphere at a coordinate radius $\xi_\ph$
and comoving time $\eta$, the photon must leave the center of the fireball
at time $\eta_1=\eta e^{-\xi_\ph}$, by equation (\ref{geod}). The photon
arrives at a radius $\xi^\prime$ at time $\eta^\prime = \eta_1 e^{\xi^\prime}
= \eta e^{\xi^\prime-\xi_\ph}$. The total energy of radiation contained in the
spheres that have been passed by the photon, evaluated along the geodesics
of the photon, is $\int e_\rr dV_\com \propto\int\eta^{\prime -1} \sinh^2
\xi^\prime d\xi^\prime\propto \int e^{-\xi^\prime} \sinh^2\xi^\prime d
\xi^\prime$. The total number of photons contained in the spheres is $\int 
n_\rr dV_\com \propto \int\sinh^2 \xi^\prime d\xi^\prime$.

Let us define
\begin{eqnarray}
	\overline{\eta} \equiv \frac{1}{2}\left(\eta+\eta_1\right) \;,
	\hspace{1cm}
	\overline{\xi} \equiv \ln\left[\frac{1}{2}\left(1+e^{\xi_\ph}\right)
		\right] \;,
	\label{eta_bar}
\end{eqnarray}
on the geodesics of the photon. By equations (\ref{rph_eq}) and 
(\ref{tau_eta}), the optical depth at $\left\{\overline{\eta},\overline{\xi}\right\}$ is
\begin{eqnarray}
	\overline{\tau} = \frac{\eta_0^2}{\eta^2}\left(1-\frac{\eta^2}
		{\eta_0^2}\right)\left[\frac{4 e^{2\xi_R}}{\left(1+
		e^{\xi_R}\sqrt{1-\eta^2/\eta_0^2}\right)^2}-1\right] \;.
	\label{tau_bar}
\end{eqnarray}

The fraction of the energy of radiation contained in the spheres that have 
been passed by the photon as it arrives at $\left\{\overline{\eta},
\overline{\xi}\right\}$, to the total energy of radiation contained in the 
spheres that have been passed by the photon as it arrives at the photosphere, 
is
\begin{eqnarray}
	f_{E_\rr} &=& \frac{\int_0^{\overline{\xi}} e^{-\xi^\prime} \sinh^2
		\xi^\prime d\xi^\prime}{\int_0^{\xi_\ph} e^{-\xi^\prime} 
		\sinh^2\xi^\prime d\xi^\prime} \nonumber \\
		&=& \frac{3 e^{\overline{\xi}} + 6 e^{-\overline{\xi}}
		-e^{-3\overline{\xi}} -8}{3 e^{\xi_\ph} + 6 e^{-\xi_\ph}
		-e^{-3\xi_\ph} -8} \;.
	\label{fr1}
\end{eqnarray}

Similarly, the fraction of the number of photons contained in the spheres 
that have been passed by the photon as it arrives at $\left\{\overline{\eta},
\overline{\xi}\right\}$, to the total number of photons contained in the 
spheres that have been passed by the photon as it arrives at the photosphere, 
is
\begin{eqnarray}
	f_{N_\rr} = \frac{\int_0^{\overline{\xi}} \sinh^2\xi^\prime 
		d\xi^\prime}{\int_0^{\xi_\ph}\sinh^2\xi^\prime d
		\xi^\prime} 
		= \frac{\sinh 2\overline{\xi} -2\overline{\xi}}
		{\sinh 2\xi_\ph -2\xi_\ph} \;.
	\label{fr2}
\end{eqnarray}

In Fig.~\ref{energy_tau}, we plot $f_{E_\rr}$ and $f_{N_\rr}$ as functions of 
$\eta$ on the photosphere. The solid curves are for $\Gamma=100$. The dotted 
curves are for $\Gamma=1000$. The values of $1-f_{E_\rr}$ and $1-f_{N_\rr}$ 
give, respectively, the fraction of the total radiation energy and the 
fraction of the total number of photons contained in the spheres that have
been passed by the photon on its last half journey to the photosphere. When 
$\eta\la 0.999\eta_\ph$, these fractions are respectively about $50$ percent 
and $75$ percent. As $\eta\rightarrow\eta_\ph$, the fractions approach $87.5$ 
percent.

Inside the photosphere, the radiation has a thermal spectrum with a comoving 
temperature $T\propto\eta^{-1}$. The effective temperature of the radiation
emitted by the photosphere, measured in the particle's comoving frame at the 
photosphere, can be approximated by
\begin{eqnarray}
	T_\eff = T \overline{\tau}^{-1/4} \;, \label{tem_eff}
\end{eqnarray}
where $T$ is evaluated on the photosphere. Submitting equation (\ref{tau_bar})
into equation (\ref{tem_eff}), and letting $T=T_0 \eta_0/\eta$ where $T_0$ is 
the comoving temperature of the radiation at $\eta=\eta_0$, we get
\begin{eqnarray}
	T_\eff &=& T_0 \left(\frac{\eta}{\eta_0}\right)^{-1/2}
		\left(1-\frac{\eta^2}{\eta_0^2}\right)^{-1/4} \nonumber\\
		&&\times \left[\frac{4 e^{2\xi_R}}{\left(1+e^{\xi_R}
		\sqrt{1-\eta^2/\eta_0^2}\right)^2}-1\right]^{-1/4} \;.
	\label{tem_eff2}
\end{eqnarray}

At $\eta=\eta_\ph$ (where $\xi_\ph=0$), we have $\overline{\tau} =1$ and 
$T_\eff = T_0 \left[1-\exp(-2\xi_R)\right]^{-1/2}$. When $\eta\ll \eta_0$,
we have $\overline{\tau}\approx 3 \eta_0^2/\eta^2$ and $T_\eff \approx
3^{-1/4} T_0 (\eta/\eta_0)^{-1/2}$.

\section{The Quasi-Thermal Precursor}
\label{precursor}

In Section \ref{photo}, we have seen that the total optical depth for a 
photon emitted from the center of the fireball is $\propto \eta^{-2}$, where
$\eta$ is the comoving time when the photon is emitted (eq. \ref{tau_eta}). 
When $\eta>\eta_\ph$, where $\eta_\ph$ is defined by equation (\ref{eta_ph}), 
the whole fireball is transparent to photons. When $\eta<\eta_\ph$, the 
fireball has a photosphere, within which the plasma is optically thick.
The emission from the photosphere has a quasi-thermal spectrum. As we will
see, since the emission has a finite amount of energy, it produces a transient
event corresponding to the precursor of a GRB. 

The photosphere is determined by equation (\ref{rph_eq}). From this section 
onward, we assume the ultra-relativistic limit $\Gamma\gg 1$. Then, we have 
$\eta_\ph\approx \eta_0$ and $e^{\xi_R}\approx 2\Gamma$. Using the coordinates 
$\{t,r\}$, the photosphere hypersurface is determined by the functions of 
parameter $\eta$
\begin{eqnarray}
	r_\ph &=& \Gamma c\eta\left(\sqrt{1-\eta^2/\eta_0^2}
		-\frac{1}{4\Gamma^2 \sqrt{1-\eta^2/\eta_0^2}}\right) \;,
	\label{rph_eta} \\
	t_\ph &=& \Gamma \eta\left(\sqrt{1-\eta^2/\eta_0^2}
		+\frac{1}{4\Gamma^2 \sqrt{1-\eta^2/\eta_0^2}}\right) \;,
	\label{tph_eta}
\end{eqnarray}
where $0<\eta<\eta_\ph$.

The velocity of a particle on the photosphere is $\beta_\ph c = r_\ph/t_\ph$. 
The Lorentz factor is $\gamma_\ph = t_\ph/\eta$.

The specific flux density of the radiation emitted by the photosphere as
measured by a remote observer, whose distance to the fireball is much larger
than the radius of the fireball, is
\begin{eqnarray}
	F_{E_\ob} = \int I_{E_\ob} d\Omega_\ob \;, \label{feo}
\end{eqnarray}
where $E_\ob$ is the observed photon energy, $I_{E_\ob}$ is the observed
specific intensity of the radiation, and $d\Omega_\ob$ is the element of the
solid angle subtended by the image of the photosphere on the observer's sky.

Because of the fact that $I_{E_\lo}/E_\lo^3$ is invariant along the path of 
a photon, where $E_\lo$ is the energy measured by any local observer on the 
path (Misner, Thorne \& Wheeler 1973), we have $I_{E_\ob} = I_{E_\emm} E_\ob^3
/E_\emm^3$, where $E_\emm$ is the energy of the photon at its point of 
emission on the photosphere as measured by an observer comoving with the 
particle emitting the photon, and $I_{E_{\rm em}}$ is the specific intensity 
measured by that observer. Then, equation~(\ref{feo}) can be rewritten as
\begin{eqnarray}
	F_{E_\ob} = \int g^3 I_{E_\emm} d\Omega_\ob \;,
	\label{feo2}
\end{eqnarray}
where 
\begin{eqnarray}
	g\equiv \frac{E_\ob}{E_\emm} = \frac{1}{\gamma_\ph\left(1-\beta_\ph
		\cos\theta\right)} 
	\label{red_shift}
\end{eqnarray}
is the Doppler factor of the photon, where $\theta$ is the angle between the
photon's wave vector and the velocity of the particle emitting the photon
\citep{jac99}.

The local specific intensity of the blackbody radiation emitted by the 
photosphere has a Planck spectrum
\begin{eqnarray}
	I_{E_\emm} = \frac{1}{h^3c^2} \frac{2 E_\emm^3}{\exp\left(E_\emm/
		k_\bo T_\eff\right) -1} \;, \label{iem}
\end{eqnarray}
where $k_\bo$ is the Boltzmann constant, $h$ is the Planck constant, and
$T_\eff$ is the effective temperature measured in the comoving frame 
(eq. \ref{tem_eff2}). Substituting equation (\ref{iem}) into equation 
(\ref{feo2}), we get
\begin{eqnarray}
	F_{E_\ob} = \frac{2 E_\ob^3}{h^3c^2} \int \frac{d\Omega_\ob}
		{\exp\left(E_\ob/gk_\bo T_\eff \right) -1} \;,
	\label{feo3}
\end{eqnarray}
where $E_\emm = E_\ob/g$ has been used.

Without loss of generality, we assume that the observer is located on the 
polar axis of the fireball ($\theta=0$), having a distance $D$ from the 
center of the fireball. At time $t_\ph$, a photon is emitted by a particle 
on the photosphere. The velocity of the particle has an angle $\theta$ from 
the polar axis. If the photon is emitted in the direction of the observer,
it will arrive at the observer at time 
\begin{eqnarray}
	t_\ob = t_\ph - \frac{r_\ph}{c}\cos\theta \;,  \label{t_obs}
\end{eqnarray}
where we have shifted the origin of the observer's time so that $t_\ob = 0$
if the photon is emitted when the fireball has a zero radius. To arrive at
the observer at the same time, photons with different polar angles on the
photosphere must be emitted at different time.

In Section \ref{photo} we have seen that the photosphere hypersurface becomes
spacelike when $\eta>\eta_0/\sqrt{2}$. That is, after $\eta=\eta_0/\sqrt{2}$, 
the photosphere shrinks superluminally. At $\eta=\eta_0/\sqrt{2}$, we have 
$\gamma_\ph = t_\ph/\eta \approx \Gamma/\sqrt{2}$, $r_\ph \approx c t_\ph = 
\Gamma c\eta_0/2$. and $t_\ph - r_\ph/c\approx \eta_0/2\Gamma$. As we will
see, because of the relativistic beaming effect, the dominant contribution
to the observed radiation comes from a very small region on the photosphere 
around $\theta=0$. Hence, by equation (\ref{t_obs}), $\eta=\eta_0/\sqrt{2}$ 
corresponds to $t_\ob \approx \eta_0/2\Gamma$, and
\begin{eqnarray}
	t_0 \equiv \Gamma^{-1}\eta_0= \left(\frac{3\kappa E_\k}{8\pi 
		c^4\Gamma^5}\right)^{1/2}
	\label{t0}
\end{eqnarray}
is a critical time parameter of the model.

By equations (\ref{rph_eta}) and (\ref{tph_eta}), equation (\ref{t_obs})
can be transformed to a cubic equation
\begin{eqnarray}
	x^3 - \left(2\mu\sec^2\frac{\theta}{2}\right) \left(x^2+1\right) + 
		\left(1+4\Gamma^2\tan^2\frac{\theta}{2}\right) x = 0 \;,
	\label{cub}
\end{eqnarray}
where
\begin{eqnarray}
	x\equiv\frac{\eta}{\sqrt{\eta_0^2-\eta^2}} \;, \hspace{1cm}
		\mu \equiv \frac{t_\ob}{t_0} \;.
	\label{x_mu}
\end{eqnarray}
Since $0\le\eta\le\eta_\ph$, we have $0\le x\le x_\ph$, where
\begin{eqnarray}
	x_\ph\equiv\frac{\eta_\ph}{\sqrt{\eta_0^2-\eta_\ph^2}}
		= \left(e^{2\xi_R}-1\right)^{1/2}\approx 2\Gamma \;.
	\label{x_ph}
\end{eqnarray}

Equation (\ref{cub}), which can be solved analytically, has up to three real 
roots of $x$ in the range of $0\le x\le x_\ph$, depending on the values of 
$\theta$, $\mu$, and $\Gamma$. From these roots, we can calculate the 
corresponding $\eta$ by the inverse of the first equation in (\ref{x_mu}). 
Hence, we can get the comoving time when the photon is emitted from the
photosphere
\begin{eqnarray}
	\eta_\emm = \eta_\emm\left(\mu,\theta\right) \;, \label{eta_emm}
\end{eqnarray}
for a given $\Gamma\gg 1$.

The element of the solid angle, after integration over the azimuthal angle
$\phi$, can be written as a function of $\theta$ and $t_\ob$
\begin{eqnarray}
	d\Omega_\ob = \frac{2\pi r_\ph^2}{D^2} \sin\theta\cos\theta 
		d\theta \;.
	\label{solid}
\end{eqnarray}

As usual, the azimuthal angle $\phi$ spans a range of $0$--$2\pi$. However,
the polar angle $\theta$ that contributes to the emissions received by the 
remote observer spans only a range of $0$--$\theta_\m$, where
\citep{ree66,bia01}
\begin{eqnarray}
	\theta_\m \equiv \arccos\beta_\ph =
		\arcsin\frac{1}{\gamma_\ph} <\frac{\pi}{2}\;.
	\label{theta_max}
\end{eqnarray}
This is caused by the following fact. Because of the relativistic beaming 
effect, the radiation emitted by a particle on the photosphere into the 
hemisphere of a solid angle $2\pi$ around the normal to the photosphere in 
the particle's comoving frame, spans only a solid angle $2\pi(1-\beta_\ph)$ 
around the normal in the rest frame. Photons emitted to the outside of this 
solid angle will hit the optically thick interior of the photosphere and be 
absorbed and hence cannot reach the observer \citep{ree66,bia01}.

Then, substituting equation (\ref{solid}) into equation (\ref{feo3}), we get
\begin{eqnarray}
	F_{E_\ob} = \frac{4\pi E_\ob^3}{h^3c^2 D^2} \int_0^{\pi/2} 
		\frac{r_\ph^2\vartheta(\theta_\m-\theta)\sin\theta\cos
		\theta d\theta}{\exp\left(E_\ob/gk_\bo T_\eff \right) -1} \;,
	\label{feo4}
\end{eqnarray}
where $\vartheta(\theta_\m-\theta)$ is a step function. That is, 
$\vartheta(\theta_\m-\theta) = 1$ if $\theta<\theta_\m$; $=0$ otherwise.

After solving for $\eta_\emm$ as a function of $t_\ob$ and $\theta$ (eq.
\ref{eta_emm}), we can evaluate $r_\ph$ and $t_\ph$ (and hence $\beta_\ph$ and
$\gamma_\ph$) as functions of $t_\ob$ and $\theta$ by equations 
(\ref{rph_eta}) and (\ref{tph_eta}), $g$ by equation (\ref{red_shift}), 
$T_\eff$ by equation (\ref{tem_eff2}), and $\theta_\m$ by equation 
(\ref{theta_max}). Then, the integral in equation (\ref{feo4}) can be 
evaluated.

We remark that, when equation (\ref{cub}) has multiple roots, the contribution
of each root to the spectrum integral in equation (\ref{feo4}) (as well as to 
the luminosity and photon rate integrals in eqs. \ref{lum_th} and 
\ref{pnum_th} below) should be summed. 

The luminosity as measured by the observer is then
\begin{eqnarray}
	L &=& 4\pi D^2 \int_0^\infty F_{E_\ob} dE_\ob \nonumber\\
		&=& 8\pi \sigma_\sb \int_0^{\pi/2} g^4 T_\eff^4 r_\ph^2
		\vartheta(\theta_\m-\theta)\sin\theta\cos\theta d\theta \;,
	\label{lum_th}
\end{eqnarray}
where $\sigma_\sb = 2\pi^5 k_\bo^4/(15h^3c^2)$ is the Stefan-Boltzmann 
constant. It can be checked that when the $r_\ph$ and $T_\eff$ are constants 
and $g=1$, equation (\ref{lum_th}) gives the standard result for the 
luminosity of a spherical blackbody.

The lightcurve of a GRB or its precursor is generally not expressed in
luminosity. Instead, it is expressed in the photon rate, i.e., the number of 
photons received by the observer per unit time. By equation (\ref{feo4}), the 
photon rate, which we denote by $N$, is
\begin{eqnarray}
	N &=& 4\pi D^2 \int_0^\infty F_{E_\ob} d\ln E_\ob
		\nonumber\\
		&=& \frac{240\zeta(3)\sigma_\sb}{\pi^3 k_\bo} \nonumber\\
		&&\times \int_0^{\pi/2} g^3 T_\eff^3 r_\ph^2
		\vartheta(\theta_\m-\theta)\sin\theta\cos\theta d\theta \;,
	\label{pnum_th}
\end{eqnarray}
where the Riemann zeta function $\zeta(3)\approx 1.202$.

\section{Numerical Results}
\label{numer}

At very early time when $\eta^2/\eta_0^2\ll 1$, the photosphere expands with 
a constant velocity that is close to the speed of light, and the formulae for 
the spectrum, the luminosity, and the photon rate can be simplified 
considerably. The results in this limiting case are presented in Appendix 
\ref{const_vel}. In fact, the luminosity and the photon rate can be worked 
out analytically. It turns out that the luminosity $L$ is a constant, and the 
photon rate $N\propto \mu^{1/2}$, where the dimensionless time $\mu$ is 
defined by equation (\ref{x_mu}).

Although the spectrum cannot be worked out analytically, a remarkable feature
of the spectrum is still identified. When $t_\ob$, $E_\ob$, and $E_\ob 
F_{E_\ob}$ are scaled respectively by $t_0$ (eq. \ref{t0}), $E_{\ph,0}$ 
(eq. \ref{chi}), and $F_0$ (eq. \ref{psi}), the spectrum does not depend on 
the value of $\Gamma$, provided that $\Gamma\ga 100$ (eq. \ref{psi2}). As will 
be seen in this section, this `similarity' feature persists in the late 
evolution of the spectrum, the luminosity, and the photon rate up to a time 
$\mu\approx 10$.

From equations (\ref{red_shift}) and (\ref{t_obs}), we have $g=\Gamma\eta
/\mu\eta_0$. Then, using the variables $\chi$ defined in equation (\ref{chi})
and $x$ defined in equation (\ref{x_mu}), we can write the spectrum integral
in equation (\ref{feo4}) as
\begin{eqnarray}
	\Psi_E = \frac{45}{7\pi^4}\Gamma^2\chi^4 \int_0^{\pi/2}\frac{x^2f_1(x)
		\vartheta(\theta_\m-\theta)\sin\theta\cos\theta d\theta}
		{\exp\left[\mu\chi x^{-1/2}f_2^{1/4}(x)\right]-1} \;,
	\label{psi_feo}
\end{eqnarray}
where $\Psi_E \equiv E_\ob F_{E_\ob}/F_0$ (eq. \ref{psi}),
\begin{eqnarray}
	f_1(x) \equiv \left(\frac{1}{1+x^2}-\frac{1}{4\Gamma^2}\right)^2 \;,
	\label{f1x}
\end{eqnarray}
and
\begin{eqnarray}
	f_2(x) \equiv \frac{1}{3} \left[\frac{16\Gamma^2}{\left(1+
		2\Gamma/\sqrt{1+x^2}\right)^2} -1\right] \;.
	\label{f2x}
\end{eqnarray}

Similarly, for the luminosity and the photon rate, we have
\begin{eqnarray}
	\Psi_L = \frac{3\Gamma^2}{7\mu^4} \int_0^{\pi/2}
		x^4 f_1(x) f_2^{-1}(x) \vartheta(\theta_\m-\theta)
		\sin\theta\cos\theta d\theta \;,
	\label{lum_lum0}
\end{eqnarray}
and
\begin{eqnarray}
	\Psi_N \hspace{-0.6cm}&&= \frac{5\Gamma^2}{\left(8\sqrt{2}-2\right)
		\mu^3} \times \nonumber\\
		&& \int_0^{\pi/2}x^{7/2} f_1(x) f_2^{-3/4}(x) 
		\vartheta(\theta_\m-\theta)\sin\theta\cos\theta d\theta \;,
	\label{num_num0}
\end{eqnarray}
where $\Psi_L\equiv L/L_0$, $L_0$ is the luminosity as $\mu\rightarrow 0$ 
(eq. \ref{lum_th0}); and $\Psi_N\equiv N/N_0$, $N_0$ is defined by 
equation (\ref{pnum0}).

Equations (\ref{psi_feo}), (\ref{lum_lum0}), and (\ref{num_num0}) have the
forms that are convenient for numerical calculations.

\begin{figure}
\vspace{2pt}
\includegraphics[angle=0,scale=0.467]{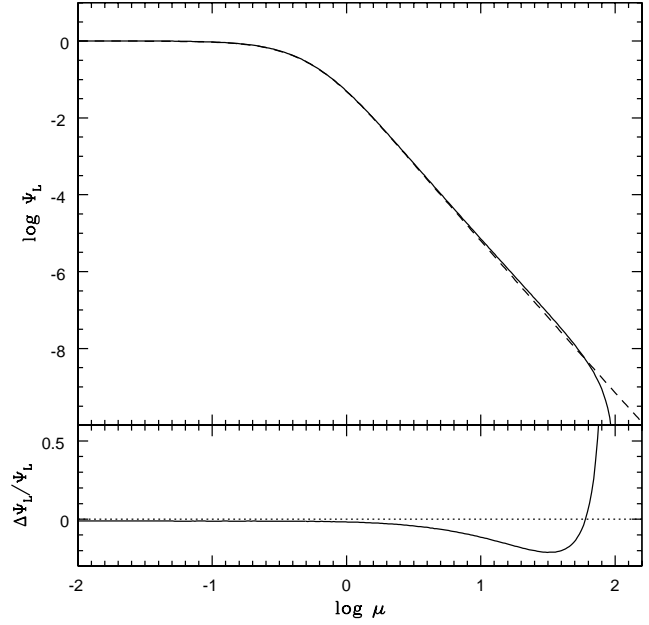}
\caption{Upper panel: the luminosity of the precursor as a function of the 
observer's time. The horizontal axis is $\mu=t_\ob/t_0$. The vertical axis is 
$\Psi_L=L/L_0$. The solid line is for $\Gamma=100$. The dashed line is for 
$\Gamma=1000$. Lower panel: the fractional difference in the luminosity: 
$\Delta\Psi_L/\Psi_L \equiv \Psi_L(\Gamma=1000)/\Psi_L(\Gamma=100) -1$. The 
dotted line marks the position of $\Delta\Psi_L=0$.
}
\label{luminosity}
\end{figure}

\begin{figure}
\vspace{2pt}
\includegraphics[angle=0,scale=0.462]{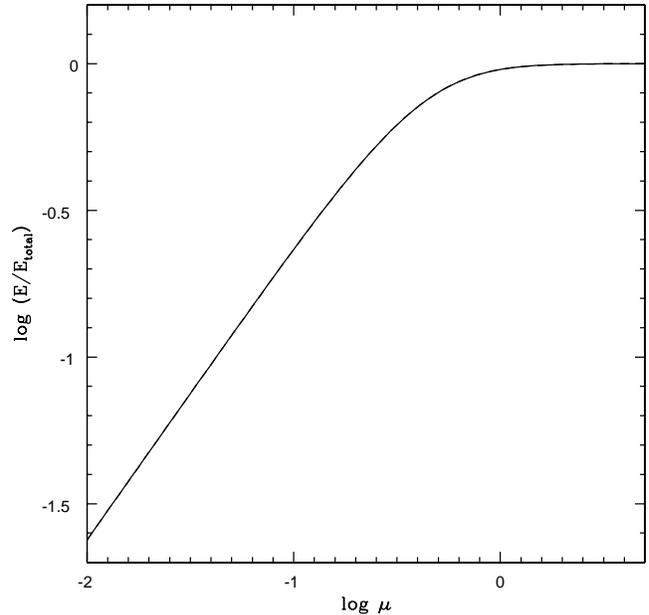}
\caption{The fraction of the accumulated energy in the total energy 
(eq. \ref{ener}). The solid line is for $\Gamma=100$. The dashed line is for 
$\Gamma=1000$ (almost indistinguishable from the solid line). 
}
\label{energy}
\end{figure}

First, we calculate the luminosity as a function of $\mu$ and $\Gamma$. The 
results are shown in Fig.~\ref{luminosity}, where the solid line is for 
$\Gamma = 100$, and the dashed line is for $\Gamma=1000$. At small $\mu$, the
solid line is almost identical to the dashed line. Quantitatively, for
$\mu<10$, the fractional difference in the two luminosities is $\la 10$ 
percent. As $\mu\rightarrow 0$, the luminosity $L$ approaches the constant 
$L_0$ (Appendix \ref{const_vel}). From $\mu\sim 0.1$, the luminosity starts 
to decay with time. At $\mu=1$, $L$ drops to $0.049L_0$. At very large $\mu$, 
the luminosity depends on the value of $\Gamma$.

The luminosity drops to zero as $\mu$ approaches $\Gamma$. This is caused by 
the following fact. On the photosphere at any comoving time $\eta$, the 
photon emitted by a particle at a larger $\theta$ reaches the observer at 
later time. By equation (\ref{t_obs}), the last photon reaches the observer 
at $t_\ob= t_\ph - r_\ph \cos\theta_\m/c = t_\ph \left(1-\beta_\ph^2\right) 
=\eta/\gamma_\ph$, since photons emitted by particles at $\theta>\theta_\m$
are invisible to the observer. Since $\gamma_\ph\ge 1$, we have $t_\ob
\le \eta \le \eta_0$, hence $\mu = t_\ob/t_0\le \Gamma$. That is, the last
photon reaches the observer at $\mu = \Gamma$.

We can define the accumulated energy of the precursor by
\begin{eqnarray}
	E\left(t_\ob\right) = \int_0^{t_\ob} L dt_\ob = E_0\int_0^\mu 
		\Psi_L d\mu \;,
	\label{ener}
\end{eqnarray}
where
\begin{eqnarray}
	E_0 \equiv L_0 t_0 = \Gamma^{-1} L_0\eta_0  \label{ener0}
\end{eqnarray}
is a characteristic total energy scale. The total energy emitted by the
photosphere is then $E_\tot=E\left(t_\ob=\Gamma t_0\right)$, which is equal to 
$0.426 E_0$ when $\Gamma=100$, and $0.421 E_0$ when $\Gamma=1000$.

\begin{figure}
\vspace{2pt}
\includegraphics[angle=0,scale=0.464]{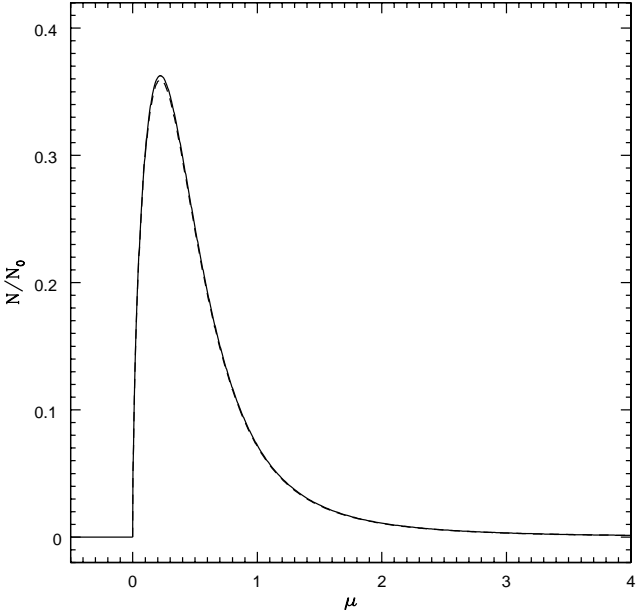}
\caption{The lightcurve of the precursor. The horizontal axis is the
observer's time in units of $t_0$. The vertical axis is the number of photons 
per unit time, in units of $N_0$. The solid curve is for $\Gamma=100$. The 
dashed curve is for $\Gamma=1000$.
}
\label{lightcurve}
\end{figure}

In Fig.~\ref{energy} we show the fraction of the accumulated energy in the
total energy. At $\mu=1$, about $96$ percent of the total energy has been 
accumulated.

The lightcurve defined by $\Psi_N=N/N_0$ as a function of $\mu$ is plotted in 
Fig.~\ref{lightcurve}, which has a smooth and FRED shape. It starts increasing
with time as $\Psi_N\propto \mu^{1/2}$, reaches a peak at $\mu = 0.221$, then 
decreases with time quasi-exponentially. The lightcurve has a long tail of 
soft photons, which lasts until $\mu=\Gamma$ (see Fig.~\ref{luminosity}). 
These soft photons originate from the spacelike part of the photosphere 
(Fig.~\ref{fireball}).

The width of the lightcurve pulse at $N=N_{\max}/4$ is $\Delta\mu = 0.895$. 
The width at $N=N_{\max}/20$ is $\Delta\mu = 1.68$. The maximum of the photon 
rate varies slightly with $\Gamma$. When $\Gamma=100$, we have $N_{\max} = 
0.363 N_0$. When $\Gamma=1000$, we have $N_{\max} = 0.359 N_0$. The variation 
in $N_{\max}$ is at a level of 1 percent.

\begin{figure}
\vspace{2pt}
\includegraphics[angle=0,scale=0.468]{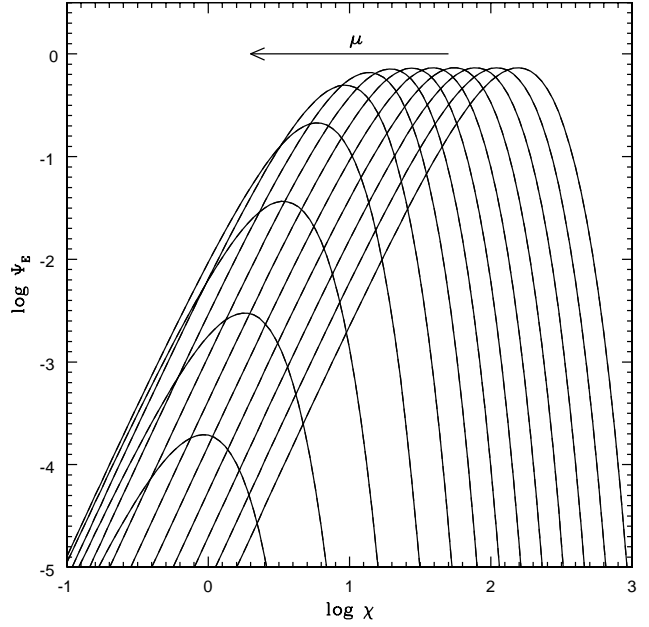}
\caption{The blackbody spectrum of the precursor. The horizontal axis is 
$\chi=E_\ob/E_{\ph,0}$, which represents the photon energy measured by the 
observer. The vertical axis is $\Psi_E = E_\ob F_{E_\ob}/F_0$, which 
represents the energy flux density measured by the observer (see the text 
for details). Different curves correspond to different observer's time, $\mu
= t_\ob/t_0 = 10^{-3}$--$10^{0.6}$ ($\Delta \log \mu = 0.3$; from right to 
left as indicated by the arrow). The solid curves are for $\Gamma=100$. 
The dashed curves (visually indistinguishable from the solid curves) are for 
$\Gamma=1000$.
}
\label{spectrum}
\end{figure}

The spectra numerically calculated by equation (\ref{psi_feo}) is plotted
in Fig.~\ref{spectrum}, for a sequence of time from $\mu = 10^{-3}$ to $\mu
= 10^{0.6}$, with the step in $\log \mu$ being $0.3$. As mentioned above,
for the range of time being of interest ($0<\mu<10$), the spectrum $\Psi_E$ 
is insensitive to the variation of $\Gamma$, provided that $\Gamma\ga 100$, 
$E_\ob$ is scaled by $E_{\ph,0}$, and $t_\ob$ is scaled by $t_0$. In 
Fig.~\ref{spectrum}, we plot the spectrum with $\Gamma=100$ with solid lines, 
and the spectrum with $\Gamma=1000$ with dashed lines. As can be seen from 
the figure, the two results are almost indistinguishable.

\begin{figure}
\vspace{2pt}
\includegraphics[angle=0,scale=0.461]{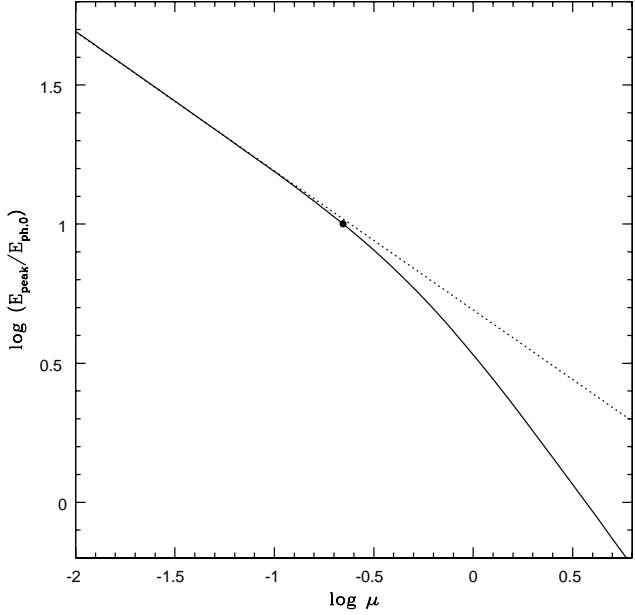}
\caption{The peak energy of the blackbody spectrum as a function of time
($\Gamma = 100$). The point marks the peak energy at the maximum of the 
lightcurve (at $\mu=0.221$). The dashed line is the peak energy calculated 
by equation (\ref{epeak}). 
}
\label{peak_energy}
\end{figure}

A critical parameter characterizing the energy of the photons is the peak
spectral energy, $E_\p$, which is defined to be the photon energy at the 
maximum of $E_\ob F_{E_\ob}$. In Fig.~\ref{peak_energy}, we plot $E_\p$ as a 
function of time. Since the results are not sensitive to the variation of 
$\Gamma$ as we have claimed by Figs.~\ref{luminosity}--\ref{spectrum}, in 
Fig.~\ref{peak_energy} (similarly in the following figures unless otherwise 
stated) we show only the plot for $\Gamma=100$. We see that, as $\mu
\rightarrow 0$, $E_\p$ approaches the asymptotic value given by (Appendix 
\ref{const_vel})
\begin{eqnarray}
	E_\p = 4.913\, E_{\ph,0} \mu^{-1/2} \;.
	\label{epeak}
\end{eqnarray}
Beyond the maximum of the lightcurve (i.e., when $\mu>0.221$), the value of 
$E_\p$ decays with time faster than that given by equation (\ref{epeak}).

\begin{figure}
\vspace{2pt}
\includegraphics[angle=0,scale=0.468]{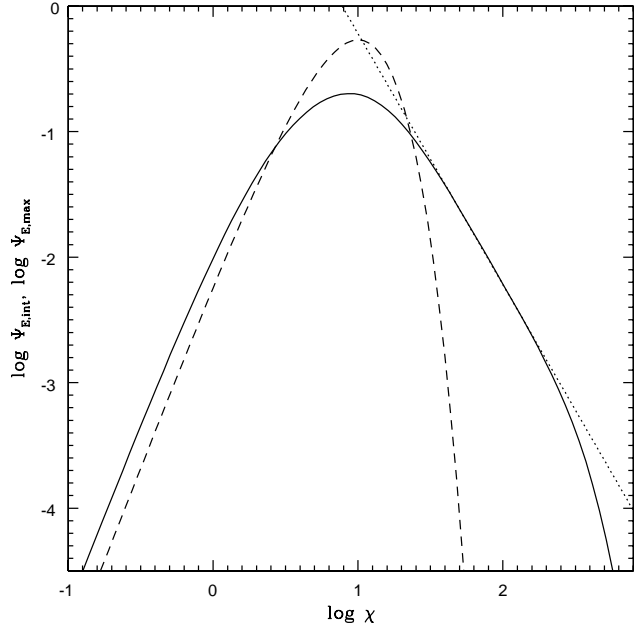}
\caption{The integrated spectrum of the precursor ($\Psi_{E,{\rm int}}$, 
integrated over the time interval of $N/N_{\max}>1/20$, eq.~\ref{psi_int}; 
solid curve), and the spectrum at the maximum of the lightcurve ($\Psi_{E,
\p}$, at $\mu=0.221$; dashed curve). The dotted line is the asymptotic 
spectrum at high energy end when the spectrum is integrated from $\mu=0$, 
$\Psi_{E,{\rm int}}\propto \chi^{-2}$ (eq.~\ref{chi_inf}). 
}
\label{spectrum_int}
\end{figure}

In Fig.~\ref{spectrum_int}, we show the spectrum at the maximum of the 
lightcurve ($\Psi_{E,\p} = \Psi_E$ at $\mu=0.221$, Fig.~\ref{lightcurve}), and 
the spectrum integrated over the time interval of $N/N_{\max}>1/20$
\begin{eqnarray}
	\Psi_{E,{\rm int}} = \int_{\mu_1}^{\mu_2} \Psi_E d\mu \;, 
	\label{psi_int}
\end{eqnarray}
where $\mu_1=0.000345$ and $\mu_2=1.68$. At the low energy end, $\Psi_{E,
{\rm int}} \propto \chi^3$, just as the instant spectrum at any moment. At
the high energy end, $\Psi_{E,{\rm int}} \propto \chi^{-2}$, if we take
$\mu_1 =0$. 

The integrated spectrum, $\Psi_{E,
{\rm int}}$, peaks at $\chi = 8.241$. The spectrum at the maximum of the 
lightcurve, $\Psi_{E,\p}$, peaks at $\chi = 10.02$. Therefore, the peak 
spectral energy at the maximum of the lightcurve is
\begin{eqnarray}
	E_{\p,\max} = 10.02\,E_{\ph,0} \;.   \label{ep_max}
\end{eqnarray}
The peak spectral energy of the integrated spectrum is
\begin{eqnarray}
	E_{\p,{\rm int}} = 8.241\,E_{\ph,0} \;.   \label{ep_int}
\end{eqnarray}

\section{Characteristic Quantities}
\label{char}

Our model has three characteristic quantities: $E_{\ph,0}$ (eq. \ref{chi}), 
a photon energy scale charactering the peak energy of the spectrum; $t_0$ (eq.
\ref{t0}), a time scale characterizing the duration and the `pulse width' of 
the lightcurve; and $E_0$ (eq. \ref{ener0}), an energy scale characterizing 
the total energy emitted by the photosphere.

The ratio $E_0/E_\k$ determines the dynamical importance of the radiation 
field to the fireball. As mentioned in Section \ref{rad}, the importance of 
the radiation field for the dynamics of the fireball is determined by the 
parameter $\lambda$ defined by equation (\ref{model_cond}), where $e_\rr$ is 
the comoving energy density of the radiation, and $\rho$ is the comoving mass 
density of the fireball. Since $e_\rr = aT^4 = a T_0^4(\eta/\eta_0)^{-4}$, 
where $T$ is the comoving temperature, and $a=4\sigma_\sb/c$ is the 
radiation density constant, by equations (\ref{vol_com_rel}), 
(\ref{mass}), and (\ref{ener_rel}) we have
\begin{eqnarray}
	\lambda \approx\frac{16\pi \sigma_\sb T_0^4 \Gamma^3 c^2\eta_0^3}
		{3E_\k}\left(\frac{\eta}{\eta_0}\right)^{-1} \;.
	\label{lam1}
\end{eqnarray}

By equations (\ref{r0}), (\ref{lum_th0}), and (\ref{ener0}), we have then
\begin{eqnarray}
	\lambda \approx \frac{6L_0\eta_0}{7\Gamma E_\k}\left(\frac{\eta}
		{\eta_0}\right)^{-1}= \frac{6\epsilon_0}{7}\left(\frac{\eta}
		{\eta_0}\right)^{-1} \;,
	\label{lam2}
\end{eqnarray}
where
\begin{eqnarray}
	\epsilon_0 \equiv \frac{E_0}{E_\k} \;.  \label{lam0}
\end{eqnarray}
Hence, at $\eta=\eta_0$, we have $\lambda = 6\epsilon_0/7$. For the model of 
a freely expanding fireball to be a good approximation, we must require that
$\epsilon_0\ll 1$.

From equations (\ref{ener0}), (\ref{chi}), (\ref{lum_th0}), and the definitions
of $t_0$ (eq. \ref{t0}) and $R_0$ (eq. \ref{r0}), we can derive that
\begin{eqnarray}
	E_{\ph,0} &=& k_\bo \left(\frac{3E_0}{14\pi ac^3 t_0^3\Gamma^2}
		\right)^{1/4} \nonumber\\
		&=& 2.073\,{\rm keV} \left(\frac{E_0}{10^{52}{\rm erg}}
		\right)^{1/4}\left(\frac{t_0}{1\,{\rm s}}\right)^{-3/4}
		\nonumber\\
		&&\times\left(\frac{\Gamma}{100}\right)^{-1/2}\;.
	\label{eph0}
\end{eqnarray}
If $E_0$, $E_{\ph,0}$, and $t_0$ can be measured, then equation (\ref{eph0})
can be used to estimate the value of the Lorentz factor $\Gamma$.

Since $\Gamma>1$, equation (\ref{eph0}) leads to a constraint on $E_0$, $t_0$,
and $E_{\ph,0}$
\begin{eqnarray}
	 E_{\ph,0} < 20.73\,{\rm keV} \left(\frac{E_0}{10^{52}{\rm erg}}
		\right)^{1/4}\left(\frac{t_0}{1\,{\rm s}}\right)^{-3/4} \;.
	\label{eph_max}
\end{eqnarray}

As we have assumed, the fireball does not contain electron-positron pairs and
the opacity in the fireball is dominated by the Thompson electron scattering
opacity. Then we have $\kappa = 0.20 (1+X)$ cm$^2$ g$^{-1}$, where $X$ is the 
mass fraction of hydrogen. It is likely that long-duration GRBs arise from the 
core-collapse of massive stars which have lost their hydrogen envelopes 
\citep{woo06}, hence it is reasonable to assume that $X=0$. Then we have 
$\kappa = 0.2$ cm$^2$ g$^{-1}$. 

Then, by equation (\ref{t0}), we have
\begin{eqnarray}
	t_0 &\approx& 1.719\,{\rm s}\, \left(\frac{\kappa}{0.2\,
		{\rm cm}^2{\rm g}^{-1}}\right)^{1/2} \left(\frac{E_\k}
		{10^{54}{\rm erg}}\right)^{1/2} \nonumber\\
		&&\times\left(\frac{\Gamma}{100}\right)^{-5/2} \;.
	\label{t0_num} 
\end{eqnarray}
By the definition of $\epsilon_0$, equation (\ref{lam0}), we have
\begin{eqnarray}
	E_0 ~=~ 10^{52}{\rm erg}\, \left(\frac{\epsilon_0}{0.01}\right)
		\left(\frac{E_\k}{10^{54}{\rm erg}}\right) \;.
	\label{ener0_num}
\end{eqnarray}
By equations (\ref{t0}), (\ref{lam0}), and (\ref{eph0}), we have
\begin{eqnarray}
	E_{\ph,0} &\approx& 1.381\, {\rm keV}\, \left(\frac{\epsilon_0}
		{0.01}\right)^{1/4} \left(\frac{\kappa}{0.2\,
		{\rm cm}^2{\rm g}^{-1}}\right)^{-3/8}  \nonumber\\
		&&\times \left(\frac{E_\k}{10^{54}{\rm erg}}\right)^{-1/8}
		\left(\frac{\Gamma}{100}\right)^{11/8} \;.
	\label{eph0_num}
\end{eqnarray}

By equations (\ref{eph0_num}) and (\ref{chi}), we have 
\begin{eqnarray}
	k_\bo T_0 &\approx& 0.018\, {\rm keV}\, \left(\frac{\epsilon_0}
		{0.01}\right)^{1/4} \left(\frac{\kappa}{0.2\,
		{\rm cm}^2{\rm g}^{-1}}\right)^{-3/8}  \nonumber\\
		&&\times \left(\frac{E_\k}{10^{54}{\rm erg}}\right)^{-1/8}
		\left(\frac{\Gamma}{100}\right)^{3/8} \;.
	\label{tem0_num}
\end{eqnarray}
The very small value of $k_\bo T_0$ justifies our assumption that the fireball
does not contain pairs.\footnote{Even if $E_{\ph,0}$ has a value as high as 
$\sim 260$ keV which is likely the case when the blackbody spectrum is 
distorted by electron scattering (Section \ref{distortion}), we have $k_\bo T_0
\sim 3$ keV and the effect of pair production is also unimportant at least 
during the stage when the majority of the precursor photons are emitted.}

By equations (\ref{t0}) and (\ref{r0}), the characteristic radius of the
fireball when the thermal precursor emission takes place is
\begin{eqnarray}
	R_0 = \Gamma^2 c t_0 &\approx& 5 \times 10^{14} {\rm cm}\, 
		\left(\frac{\kappa}{0.2\,{\rm cm}^2{\rm g}^{-1}}
		\right)^{1/2} \nonumber\\
		&&\times\left(\frac{E_\k}{10^{54}{\rm erg}}
		\right)^{1/2}\left(\frac{\Gamma}{100}\right)^{-1/2} \;.
	\label{rc0}
\end{eqnarray}
The radius of the fireball at the end of acceleration, defined by $\lambda
= 1$ ($\eta=\eta_\acc$), is
\begin{eqnarray}
	R_{\rm acc} = \frac{6\epsilon_0}{7} R_0 &\approx& 4 \times 10^{12} 
		{\rm cm}\, \left(\frac{\epsilon_0}{0.01}\right)
		\left(\frac{\kappa}{0.2\,{\rm cm}^2{\rm g}^{-1}}
		\right)^{1/2} \nonumber\\
		&&\times\left(\frac{E_\k}{10^{54}{\rm erg}}
		\right)^{1/2}\left(\frac{\Gamma}{100}\right)^{-1/2} 
		\;. \label{r_acc}
\end{eqnarray}
For the model of a freely expanding fireball to be self-consistent, we must
require that $R_{\rm acc}\ll R_0$, i.e. $\epsilon_0\ll 1$.

The peak spectral energy is related to $E_{\ph,0}$ by equations (\ref{epeak}),
(\ref{ep_max}), and (\ref{ep_int}). The duration of the precursor (defined by
the width of the lightcurve pulse at $N=N_{\max}/20$) is $\approx 1.7 t_0$. 
The total energy emitted in the precursor, is $\approx 0.42 E_0$.

Generally, the characteristic photon energy predicted by our simple model 
with the assumption of a blackbody photosphere emission is too small for 
explaining the precursors detected by the gamma-ray detectors, but is in 
agreement with the precursors detected by the X-ray detectors (see 
Introduction; for further discussions see Sec. \ref{distortion}).

\section{Discussions}
\label{discussion}

\subsection{The Effect of Jet Collimation}
\label{jet}

In our model we have assumed that the fireball is a sphere with a total solid
angle of $4\pi$. In reality, GRBs are likely to be collimated. That is, the
fireball may have a cone shape and spans a solid angle $\Omega<4\pi$. Now we 
check how this `jet collimation' effect affects our results.

Assume that the fireball spans a solid angle $\Omega=4\pi\omega$, $\omega<1$.
The observer is on the polar axis $\theta=0$, and the polar boundary of the
fireball (which is now a cone) is at $\theta=\theta_\c <\pi/2$, where 
$\theta_\c \equiv \arccos(1-2\omega)$. In addition, we assume that $\theta_\c
\gg \Gamma^{-1}$, so that the effect of `jet collimation' starts to affect 
the shape of the lightcurve only at very late time when the Lorentz factor 
of the photosphere ($\gamma_\ph$) drops to a value $\sim 1/\theta_\c$, just 
as the observation of the `jet break' signature in GRB afterglow lightcurves 
\citep[and references therein]{sar99,fri05}. Hence, we have $\Gamma^{-2}\ll 
\omega< 1$.

With jet collimation, the comoving volume of the fireball in equation
(\ref{vol_com_rel}) should be replaced by
\begin{eqnarray}
	V_\com \approx 2\pi\omega\Gamma^2 c^3\eta^3 \propto\omega \;,
\end{eqnarray}
where we have let $\gamma=\Gamma$. The kinetic energy $E_\k$ is still
related to the total mass $M$ by equation (\ref{ener_rel}), but equation
(\ref{mass_rel}) should be replaced by
\begin{eqnarray}
	M \approx \frac{\omega}{2}M_0 \Gamma^2 \;,
\end{eqnarray}
where $M_0$ is defined by equation (\ref{M0}).

The optical depth is still given by equations (\ref{tau_1}) and 
(\ref{tau_eta}), and $\eta_0$ is still defined by equation (\ref{eta0}). Then,
for the characteristic time in equation (\ref{t0}), we should replace $E_\k$ 
by $E_\k/\omega$. This is equivalent to say that equation (\ref{t0}) still 
holds if $E_\k$ is interpreted as the isotropic-equivalent kinetic energy.

The photosphere is still defined by equation (\ref{rph_eq}), or equivalently,
by equations (\ref{rph_eta}) and (\ref{tph_eta}). The effective temperature
of the photosphere is still given by equation (\ref{tem_eff2}).

Equation (\ref{cub}) is unchanged, except that $\theta$ is only allowed to
vary in the range of $0\le \theta\le\theta_\c$. 

The specific flux density measured by the observer is still given by equation
(\ref{feo4}), but the following replacement to the integral over $\theta$ 
should be made
\begin{eqnarray}
	\int_0^{\pi/2} d\theta ~\longrightarrow~ \int_0^{\theta_\c} 
		d\theta \;.
	\label{replace}
\end{eqnarray}

The luminosity defined by equation (\ref{lum_th}), after the replacement in
equation (\ref{replace}), should be interpreted as the isotropic-equivalent 
luminosity. The true luminosity of the photosphere should be $\omega L$. The 
same conclusion holds for the photon rate in equation (\ref{pnum_th}).

\begin{figure}
\vspace{2pt}
\includegraphics[angle=0,scale=0.464]{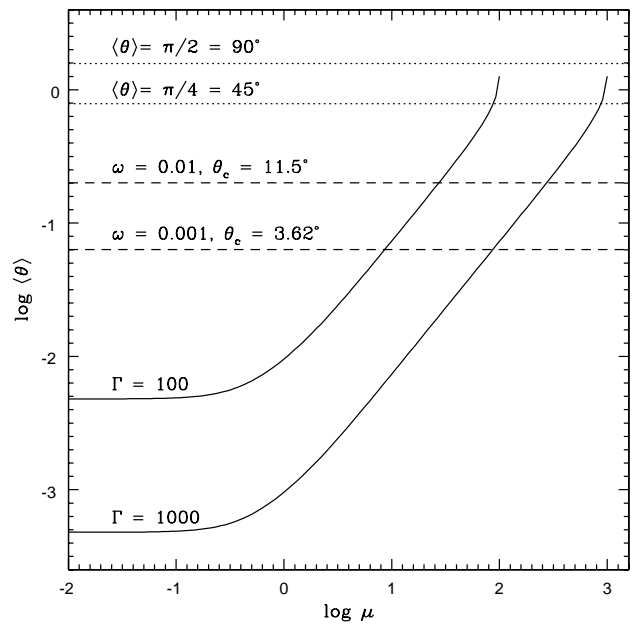}
\caption{The flux-averaged polar angle on the photosphere (solid curves), 
defined by equation (\ref{gamma_av}). The curve ends at $\mu= \Gamma$, since 
after $\mu=\Gamma$ no photons emitted by the photosphere can reach the 
observer. The dashed lines show the maximum polar angle of the cone-shape 
fireball, $\theta_\c$. The model of the spherical fireball applies only when 
the value of $\langle\theta\rangle$ is below the dashed line (i.e., when 
$\langle\theta\rangle < \theta_\c$). If the photosphere is a static and 
uniform sphere, we would have $\langle\theta\rangle = \pi/4$.
}
\label{lum_gamma}
\end{figure}

Because of the relativistic beaming effect, when $\gamma_\ph\gg 1$ the 
contribution of the photosphere to the observed luminosity, the photon rate, 
and the spectrum comes from a small region of $\theta< \gamma_\ph^{-1}\ll
\theta_\c$ on the photosphere. Hence, the effect of the boundary of the jet 
becomes important only after $\gamma_\ph$ drops to a value smaller than 
$\gamma_{\ph,{\rm br}} \approx \theta_\c^{-1}$. For $\theta_\c \gg 
\Gamma^{-1}$, we have $\gamma_{\ph,{\rm br}} \ll \Gamma$, and hence the time 
$\mu=\mu_{\rm br}$ when $\gamma_\ph$ drops to $\gamma_{\ph,{\rm br}}$ would 
be $\gg 1$.

We can define a flux-averaged polar angle on the photosphere by
\begin{eqnarray}
	\langle\theta\rangle = \frac{8\pi\sigma_\sb}{L}\int_0^{\pi/2}
		\theta g^4 T_\eff^4 r_\ph^2\vartheta\left(\theta_\m-
		\theta\right)\sin\theta\cos\theta d\theta \;,
	\label{gamma_av}
\end{eqnarray}
for the model of the spherical fireball. If the photosphere is a static and
uniform sphere (i.e., $g=1$, $T_\eff$ and $r_\ph$ are constants), we would 
have $\langle\theta\rangle = \pi/4$. If the contribution to the luminosity 
comes from a small region of $\theta\ll 1$ on the photosphere (which is the 
case when the photosphere expands relativistically), we would have $\langle
\theta\rangle\ll 1$. If the contribution to the luminosity comes from a narrow 
ring region around $\theta=\theta_0$ on the photosphere, we would expect
$\langle\theta\rangle \approx \theta_0$.

In Fig.~\ref{lum_gamma}, we show the value of $\langle\theta\rangle$ as a
function of time (solid curves). When $\mu\ll 1$, the photosphere has a
Lorentz factor $\gamma_\ph\approx\Gamma \gg 1$, the radiation received by the 
observer comes dominantly from a region of $\theta\la \Gamma^{-1}$ on the 
photosphere, hence we have $\langle\theta\rangle \sim \Gamma^{-1} \ll 1$. As
$\mu\rightarrow \mu_{\max} = \Gamma$, the radiation received by the 
observer comes dominantly from a region of $\pi/4<\theta<\pi/2$.

The dashed lines show the value of $\theta_\c$, the maximum polar angle of the 
cone-shape fireball. When $\langle\theta\rangle\ga\theta_\c$, the observer 
starts to see the boundary of the cone and the spherical model does not apply 
any more. As a result, the lightcurve starts to decay faster with time than 
that predicted by the spherical model. The transition time $\mu_\tr$ (or, the 
`jet break' time, as commonly called for the GRB afterglow) is approximately 
determined by $\langle\theta\rangle = \theta_\c$, i.e. the intersection of the 
solid curve and the dashed line. When $\theta_\c \gg \Gamma^{-1}$ as we have
assumed, the transition time $\mu_\tr$ is much larger than the duration of the
event.

Particularly, when $\mu\ll 1$, the effect of jet collimation is negligible. 
The results in Appendix \ref{const_vel} are unchanged. $E_{\ph,0}$ is still 
given by equation (\ref{chi}), $L_0$ is still given by equation 
(\ref{lum_th0}), and $N_0$ is still given by equation (\ref{pnum0}). However, 
as explained above, $L_0$ and $N_0$, as well as the $E_\k$ in the definition 
of $\eta_0$ (eq.~\ref{eta0}), should be interpreted as the 
isotropic-equivalent quantities.

Then, $E_{\ph,0}$ is still the characteristic photon energy, while $E_0 =
L_0 t_0$ is the isotropic-equivalent characteristic total energy. 
Equations (\ref{lam0})--(\ref{rc0}) are then unchanged, if the $E_\k$ and 
$E_0$ are interpreted as the isotropic-equivalent kinetic energy and the 
isotropic-equivalent characteristic total energy, respectively.

\subsection{Dependence of the Lightcurve and Spectrum on the Energy Band
of a Detector}
\label{detector} 

Every detector has a finite range of energy that the detector is sensitive to. 
For example, the BAT on \sw\, covers an energy range of 15--150 keV, while the 
BATSE on \cgro\, covers an energy range of 20--1900 keV. The observed 
lightcurve and the spectrum of a GRB precursor depends crucially on the range 
of energy of the detector.

\begin{figure}
\vspace{2pt}
\includegraphics[angle=0,scale=0.464]{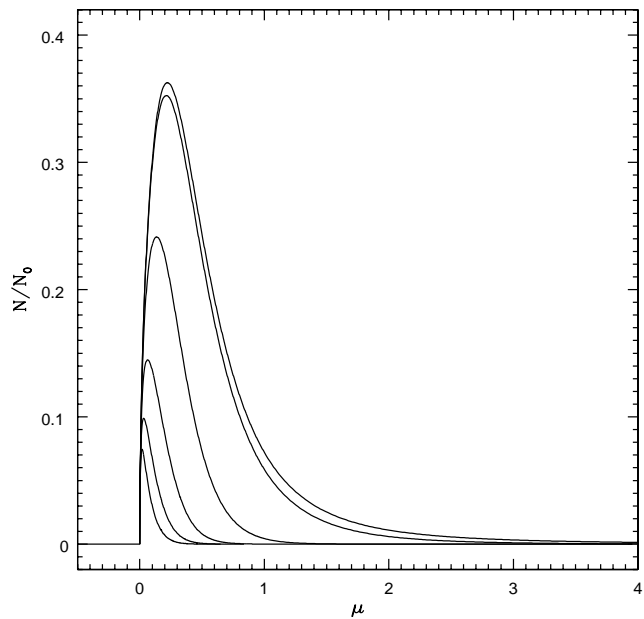}
\caption{The lightcurves detected by detectors with different lower limit
on the photon energy. Top to bottom: $E_{\min}/E_{\ph,0} = 0$, $1$, $5$,
$10$, $15$, and $20$.
}
\label{lightcurve2}
\end{figure}

We consider only the effect of the lower limit on the photon energy. For a
quasi-thermal spectrum the flux density decays quickly at the high energy
end, the upper limit on the photon energy is not important provided that it 
exceeds the peak spectral energy at the maximum of the 
lightcurve.\footnote{This could be a problem for \sw\, since the BAT on it has
a very small maximum energy.} Hence, we take the lowest energy to be 
$E_{\min}$, and allow the highest energy to be infinity. Then, the observed 
photon rate is
\begin{eqnarray}
	\Psi_N = \frac{N}{N_0} = \frac{7\pi^4}{36\left(4\sqrt{2}-1\right)
		\zeta(3)}\int_{\chi_{\min}}^{\infty} \chi^{-2} 
		\Psi_E d\chi \;,
\end{eqnarray}
where $\chi_{\min} \equiv E_{\min}/E_{\ph,0}$.

The lightcurves calculated for $\chi_{\min} = 0$, 1, 5, 10, 15, and 20 are
shown in Fig.~\ref{lightcurve2}. As $E_{\min}$ increases, the maximum of the
lightcurve shifts to earlier time, and the width of the lightcurve pulse
decreases. This is caused by the fact that the energy of the photons emitted
by the photosphere decreases with time (Figs. \ref{spectrum} and 
\ref{peak_energy}). The maximum photon rate decreases as $E_{\min}$ increases.

\begin{figure}
\vspace{2pt}
\includegraphics[angle=0,scale=0.464]{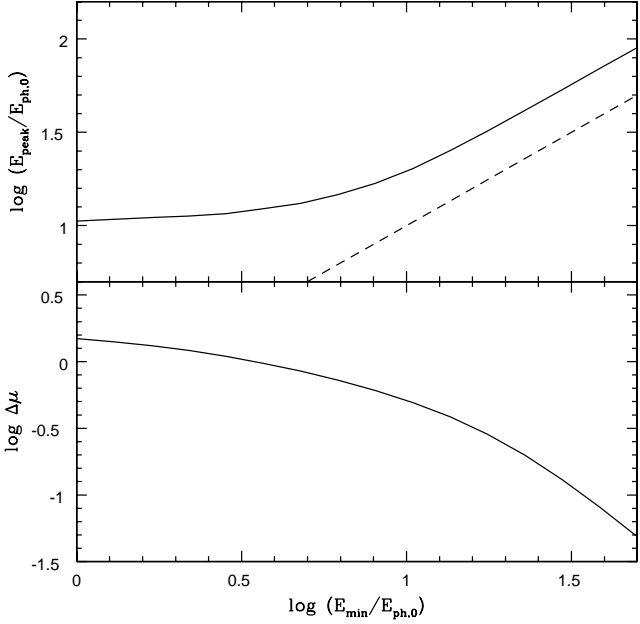}
\caption{The observed peak photon energy of the spectrum at the maximum of 
the lightcurve (upper panel) and the width of the observed lightcurve pulse 
(at $N = N_{\max}/20$; lower panel) as functions of the lower energy limit of 
the detector. The dashed line in the upper panel shows the lower energy limit 
of the detector.
}
\label{lightcurve2_peak}
\end{figure}

In Fig.~\ref{lightcurve2_peak}, we show the observed peak photon energy of the 
spectrum at the maximum of the lightcurve (estimated by eq.~\ref{epeak}) and 
the width of the observed lightcurve pulse at $N=N_{\max}/20$ as functions of 
$E_{\min}$. The peak photon energy increases with $E_{\min}$, while the width 
of the lightcurve pulse decreases with $E_{\min}$.

\subsection{Distortion of the Blackbody Spectrum by Electron Scattering}
\label{distortion}

Even if the effects of jet collimation and the finite energy band of the
detector are considered, the characteristic photon energy predicted by the
model is still significantly smaller than that of the GRB precursors detected
by gamma-ray detectors. For example, the spectrum of the precursor of 
GRB 041219a can be fitted by a blackbody plus a power law. The temperature 
of the blackbody is $\approx 46\pm 9$ keV, corresponding to a peak photon 
energy $\approx 180\pm 35$ keV (in the observer's frame; McBreen et al. 2006).
The spectrum of the precursor of GRB 030406 can be fitted by a blackbody with 
temperature $\approx 106\pm 13$ keV, corresponding to a peak photon energy 
$\approx 416\pm 51$ keV (in the observer's frame; Marcinkowski et al. 2006).
However, the peak photon energy of the integrated spectrum predicted by our 
model is typically $\approx 8 E_{\ph,0}\approx 10$ keV, which appears to be 
consistent with that of the soft X-ray precursors detected by {\em Ginga} 
\citep{mur91,mur92} and WATCH/{\em GRANAT} \citep{saz98}. 

In our model we have assumed that the emission from the photosphere is a 
perfect blackbody. However, when electron scattering dominates the opacity, 
thermal equilibrium exists only when $\tau_{\rm ff}\tau_\th>1$, where 
$\tau_{\rm ff}$ is the free-free optical depth, and $\tau_\th$ is the 
Thompson optical depth \citep{fel72,sha72,sha73,nov73,sun74}. At the surface 
of the fireball, the spectrum of the outgoing radiation is distorted. The 
energy flux measured in the comoving frame is \citep{nov73}
\begin{eqnarray}
	Q = 1.54\times 10^{-4} n_e^{1/2} T_\eff^{9/4} \label{qq}
\end{eqnarray}
in cgs units, where $n_e$ is the number density of electrons, and $T_\eff$
is the effective temperature. Equation (\ref{qq}) is in contrast to the
standard formula $Q = \sigma_\sb T_\eff^4$ for the blackbody.

The characteristic radius for the emission from the photosphere is the $R_0$ 
given by equation (\ref{rc0}). Let us define the effective temperature of the 
standard blackbody spectrum in the comoving frame, $T_{\eff,0}$, by $L_0 = 
4\pi\sigma_\sb T_{\eff,0}^4 R_0^2 \Gamma^2$, where $L_0$ is related to the 
characteristic total energy $E_0$ by equation (\ref{ener0}). Similarly, we
can define the effective temperature of the distorted blackbody spectrum, 
$T_{\eff,1}$, by $L_0 = 4\pi Q R_0^2 \Gamma^2$, where $Q$ is given by 
equation (\ref{qq}). Then, we find that
\begin{eqnarray}
	\frac{T_{\eff,1}}{T_{\eff,0}} \approx 23.5 \left(\frac{\epsilon_0}
		{0.01}\right)^{7/36} \left(\frac{E_\k}{10^{54}{\rm erg}}
		\right)^{1/72}\left(\frac{\Gamma}{100}\right)^{-1/24} ,
	\label{tem_r}
\end{eqnarray}
where we have assumed that $\kappa = 0.2$ cm$^2$ g$^{-1}$ and $n_e = \rho/
2m_{\rm H}$, where $m_{\rm H}$ is the atomic mass unit.

If we take $E_\p\approx 8 E_{\ph,0}$, then equations (\ref{eph0_num}) and 
(\ref{tem_r}) imply that
\begin{eqnarray}
	E_\p &\approx& 259\, {\rm keV}\, \left(\frac{\lambda_0}
		{0.01}\right)^{4/9} \left(\frac{E_\k}{10^{54}
		{\rm erg}}\right)^{-1/9} \nonumber\\
		&&\times\left(\frac{\Gamma}{100}\right)^{4/3} \;,
	\label{epeak_es}
\end{eqnarray}
for the distorted blackbody spectrum. The peak spectral energy given by 
equation (\ref{epeak_es}), which is defined in the observer's frame, appears 
to be in agreement with that of the precursors detected by gamma-ray 
detectors.

Another effect that may help shifting up the photon energy is that the 
blackbody emission from the photosphere may be nonthermalized by some
scattering process and have a nonthermal spectrum. For example, 
\citet{tho94} has shown that the scattering of the adiabatically cooled thermal
photons by the magnetohydrodynamic (MHD) turbulence in the photosphere can 
boost the thermal photons up to a larger peak energy and lead to a spectrum 
that have a broken power-law or a Band function \citep{ban93} shape.

\section{Summary and Conclusions}
\label{concl}

We have studied a simple model for the precursor event of a GRB, where the 
precursor is produced by the emission from the early photosphere of a 
fireball. The fireball is assumed to be homogeneous, spherical, and expand 
freely and relativistically. The fireball contains a constant amount of rest 
mass, and has a constant total kinetic energy. It also contains a radiation 
field, which drives the acceleration of the fireball expansion at the very 
early stage, but simply cools down with the expansion at the late 
free expansion stage. These assumptions make our model different from that 
has been adopted in the literature for studying the photosphere emission of 
a GRB, which usually assumes a steady wind with mass and energy injection at 
the center \citep{dai02,mes00,mes02,ram05,ree05,pee06,pee07,tho07}.

The dynamics of the fireball is the same as that of a Milne universe, except 
that the fireball has a finite size and contains a finite amount of mass and 
energy. In the same way as the Universe, at very early time the fireball is 
opaque to the photons contained in it, but later it becomes transparent 
(Fig.~\ref{fireball}). The photosphere is defined to be the hypersurface where
the total optical depth is unity. If the fireball had an infinitely large 
radius, then as in the Milne universe, the photosphere were just a spacelike 
hyperbola. However, because of the fact that the fireball has a finite radius,
the hypersurface of the photosphere is timelike when $\eta<\eta_0/\sqrt{2}$. 
It is the emission from this timelike part that makes the dominant 
contribution to the precursor emission detected by a remote observer, since 
this timelike surface moves towards the observer with a relativistic speed. 
However, the spacelike part of the photosphere produces a long tail of 
soft emission.

The luminosity, the photon rate (the lightcurve), and the blackbody spectrum 
of the precursor are calculated. The results are presented in 
Figs.~\ref{luminosity}--\ref{spectrum_int}. A remarkable feature of the
results is that, when the quantities are properly scaled and the time is not 
too large, the results are almost invariant with respect to the Lorentz
factor of the fireball. That is, if the observer time is in units of $t_0 = 
\eta_0/\Gamma$ (eq.~\ref{t0}), the luminosity is in units of $L_0$ 
(eq.~\ref{lum_th0}), the photon rate is in units of $N_0$ (eq.~\ref{pnum0}), 
and the spectrum $E_\ob F_{E_\ob}$ is in units of $F_0$ (eq.~\ref{psi}), 
then the luminosity, the photon rate, and the spectrum are insensitive to the 
variation of $\Gamma$ for $\mu = t_\ob/t_0 \la 10$, provided that $\Gamma
\ga 100$. At very late time, the results depend on the value of $\Gamma$.

Thus, a critical time scale for the precursor is $t_0$ (eqs.~\ref{t0} and 
\ref{t0_num}), which is $\sim 2$~s for typical parameters. The luminosity
starts with a constant $L = L_0$, and drops to $L\approx 0.05 L_0$ at $t_\ob
= t_0$. The total energy emitted by the photosphere is $\approx 0.42 L_0 
t_0$. About 96 percent of the total emitted energy has been accumulated
at $t_\ob = t_0$.

The lightcurve (defined by the photon rate) has a smooth and FRED shape 
(Fig.~\ref{lightcurve}). It begins as $N\propto t_\ob^{1/2}$, peaks at
$t_\ob \approx 0.22 t_0$, and then decays quasi-exponentially. The width of
the lightcurve pulse at $N = N_{\max}/4$ ($N_{\max}$ is the maximum photon
rate) is $\Delta t_\ob \approx 0.9 t_0$. The width at $N = N_{\max}/20$ is 
$\Delta t_\ob \approx 1.7 t_0$. 

The blackbody spectrum at different moment of observation is shown in 
Fig.~\ref{spectrum}. The peak spectral energy $E_\p$, defined to be the photon
energy at the maximum of $E_\ob F_{E_\ob}$, decays with time 
(Fig.~\ref{peak_energy}). Before the maximum of the lightcurve ($t_\ob< 0.22
t_0$), the peak spectral energy evolves according to, approximately, $E_\p
\propto t_{\ob}^{-1/2}$ (eq.~\ref{epeak}). After the maximum, the peak 
spectral energy decays with time with a rate faster than that given by 
equation (\ref{epeak}) (Fig.~\ref{peak_energy}). At the maximum of the 
lightcurve ($t_\ob \approx 0.22 t_0$), the spectrum peaks at $E_\p = 10.0 
E_{\ph,0}$, where $E_{\ph,0}$ is the characteristic photon energy defined by 
equation (\ref{chi}). The spectrum integrated over the time interval of 
$N>N_{\max}/20$ peaks at $E_\p = 8.24 E_{\ph,0}$ (Fig.~\ref{spectrum_int}).

The characteristic quantities describing the precursor, which include the
characteristic time, the characteristic total energy, and the characteristic
photon energy are summarized in equations (\ref{t0_num})--(\ref{eph0_num}). 
For typical parameters, the characteristic time is shorter than the duration 
of a typical long GRB (the main burst), and the characteristic photon energy 
is in the X-ray band. Because of the thermal nature of the spectrum, at least 
in the high energy band the total energy is expected to be smaller than the 
main burst. A constraint on these three characteristic quantities is given 
by equation (\ref{eph_max}), which does not depend on the value of the opacity
in the fireball.

The characteristic radius of the thermal emission, $R_0$, is given by equation 
(\ref{rc0}), which is $10^{14}$--$10^{15}$ cm for typical parameters. The 
internal shock must not occur until the fireball becomes optically thin, since 
otherwise it would be difficult to explain the nonthermal spectrum of the GRB. 
Hence, $R_\delta$---the radius where the internal shock occurs---must be 
$>R_0$. The time separation between the precursor and the main burst is then 
$t_{\rm sep} \sim R_\delta/2\Gamma^2 c$. 

The main burst produced by the internal shock has a duration $T\approx \Delta 
/c$, where $\Delta$ is the width (measured in the rest frame of the GRB) of 
the outflow that contributes to the prompt gamma-ray emission \citep{pir99}.
In our model, the fireball has an expansion velocity $v\propto r$, and the 
majority of mass and energy is confined in a thin layer at the outer boundary
of the fireball (Sec. \ref{kine}). The thickness of the layer---which
contains $50$ percent of total mass and $65$ percent of total kinetic 
energy---is given by equation (\ref{delta_r}). Letting $R=R_\delta$, we have
then $\Delta\approx R_\delta/2\Gamma^2$ and $T \approx R_\delta/2\Gamma^2 c 
\sim t_{\rm sep}$. This relation between the duration of the main burst and 
the time separation between the main burst and the precursor appears to be 
consistent with the observation (see Introduction).

We have assumed a spherical fireball. If the GRB outflow is
collimated and the resulted fireball has a configuration of a cone spanning 
a solid angle $4\pi\omega$ that satisfies $\Gamma^{-2}\ll\omega <1$, the 
results are still valid until a time is reached when the flux-averaged polar 
angle of the photosphere increases to a value $\ga\theta_\c = \arccos(1-
2\omega)$ (Fig.~\ref{lum_gamma}). However, in the results, $L$ and $N$ should 
be interpreted as the isotropic-equivalent luminosity and the 
isotropic-equivalent photon rate respectively. Equations 
(\ref{lam0})--(\ref{rc0}) still hold, but $E_\k$ and $E_0$ should be 
interpreted as the isotropic-equivalent kinetic energy and the 
isotropic-equivalent characteristic total energy of radiation, respectively. 
This fact is caused by the relativistic beaming effect, which implies that 
when the photosphere expands relativistically the dominant contribution to 
the spectrum and the luminosity observed by a remote observer comes from a 
small region on the photosphere that spans a solid angle $\sim \Gamma^{-2}$.

The dependence of the observed lightcurve and spectrum on the finite energy
range of a detector has also been discussed (Sec. \ref{detector}). Overall,
as the lower limit of the detector energy increases, the duration of the 
lightcurve decreases and the maximum of the lightcurve shifts to earlier time
(Figs.~\ref{lightcurve2} and \ref{lightcurve2_peak}). The peak photon energy 
of the spectrum increases with the lower energy bound of the detector
(Fig.~\ref{lightcurve2_peak}).

The photon energy predicted by our simple model seems to be consistent with 
the X-ray precursors of GRBs detected by {\em Ginga} and WATCH/{\em GRANAT} 
\citep{mur91,mur92,saz98}, but is too small for the precursors detected by 
gamma-ray detectors \citep{cen06,mar06,mcb06,rom06,pag07}. However, in our 
model we have assumed that the emission from the photosphere has a spectrum 
that is perfectly blackbody. When electron scattering dominates the opacity, 
the blackbody spectrum is distorted \citep{fel72,sha72,sha73,nov73,sun74}. We 
have estimated this effect by adopting a modified flux equation in equation 
(\ref{qq}) and found that the distorted spectrum could have a peak spectral 
energy of $\sim 260$ keV, which is consistent with the hard spectrum of the 
observed gamma-ray precursors (Sec. \ref{distortion}). Scattering of the 
soft blackbody photons by e.g. the magnetic turbulence in the fireball may
also lead to a nonthermal spectrum \citep{tho94}. A detailed study on 
these issues is left for our future work.

A strong assumption in our model is that there is no continuous injection of
energy and mass, and hence the total mass and the kinetic energy of the 
fireball are conserved. A big benefit from this simplified assumption is that
the shape of the lightcurve and the duration of the emission can be predicted,
in contrast to the steady wind model where the lightcurve and the duration 
depend on the rate of energy and mass injection as a function of time. 
On the physics side, this type of models with `instant release of energy' may
be relevant to the precursors that have smooth and FRED shape lightcurves. 
More complicated models with continuous injection of mass and energy may be 
relevant to the precursors that have complex lightcurves.

Although the X-Ray Telescope (XRT) on \sw\, has discovered that flares are 
common in early X-ray afterglows of GRBs and it seems that the most likely 
cause of the flares is the late-time activity of the GRB central engine 
\citep[and references therein]{bur07}, the time separation between two 
adjacent flares are usually much longer than the duration of the precursor 
event predicted by our model 
and hence the assumption of no continuous energy and mass injection remains 
valid. Of course the fireballs or expanding shells related to the late-time 
activity would also have photosphere emissions. However, the late-time 
photosphere emissions would overlap with the prompt main GRB emissions or the 
afterglows \citep{mes00,mes02,ram05,ree05,pee06,tho07} and hence are hard to 
detect [see, however, \citet{ryd04,ryd05}; and \citet{pee07}].

In summary, our model implies the existence of a quasi-thermal precursor of
a GRB. The precursor is a remnant of the thermal radiation contained in the
fireball during its initial acceleration phase, just like that the CMB is
a remnant of the radiation in the Big Bang. The precursor emission has a 
quasi-thermal spectrum, and a smooth and FRED shape lightcurve with a duration 
$\sim 1$--$5$~s. If the distortion of the blackbody spectrum by the electron 
scattering process is not important, the radiation observed by a remote 
observer is in the X-ray band with a peak photon energy $\sim 1$--$10$~keV. 
If the distortion by electron scattering is important, the peak photon energy 
could be of several hundred keV and hence be in the gamma-ray band. Under
some conditions the soft thermal photons may also be nonthermalized and
have a nonthermal spectrum. Although in reality the situation may be much 
more complicated, our simple model may provide a reasonable interpretation 
for at least a class of GRB precursors---those having smooth and FRED-shape 
lightcurves and quasi-thermal spectra.

\section*{Acknowledgments}
I am grateful to D. Giannios, F. Meyer, and H. Spruit for helpful 
conversations on gamma-ray bursts and their precursors, and K. Page, P. 
Romano, and R. Ruffini for useful communications about their works.
I thank the anonymous referee for a very helpful report which has led
improvements to the paper.

\appendix

\section{Photosphere with a Constant Expansion Velocity}
\label{const_vel}

When $\eta^2/\eta_0^2\ll 1$, the photosphere defined by equation (\ref{rph_eq})
has a radius $\xi_\ph\approx\xi_R$, constant $\gamma_\ph = \cosh\xi_\ph 
\approx \Gamma\gg 1$, and constant $\theta_\m \approx 1/\Gamma\ll 1$. That is, 
the photosphere expands with a constant and ultra-relativistic velocity. In 
this Appendix, we derive the spectrum, the luminosity, and the photon rate of 
a photosphere in this limiting case.

When $\eta^2/\eta_0^2\ll 1$, we have $t_\ph \approx \Gamma\eta$, $r_\ph\approx
R\approx \Gamma c\eta$, and $\beta_\ph \approx \beta_R \approx 1-1/2\Gamma^2$. 
Then, by equation (\ref{red_shift}), we have
\begin{eqnarray}
	g \approx \frac{1}{\Gamma\left(1-\beta_\ph\cos\theta\right)} \;.
	\label{g_a}
\end{eqnarray}
By equations (\ref{red_shift}) and (\ref{t_obs}), we have $t_\ob = \eta/g$ and
hence
\begin{eqnarray}
	\frac{\eta}{\eta_0} = \frac{\mu g}{\Gamma} \;.  \label{mu_a}
\end{eqnarray}

Then, by equations (\ref{tem_eff2}) and (\ref{mu_a}), we have
\begin{eqnarray}
	T_\eff \approx 3^{-1/4} T_0 \left(\frac{\eta}{\eta_0}\right)^{-1/2} \;,
\end{eqnarray}
and
\begin{eqnarray}
	\frac{E_\ob}{g k_\bo T_\eff} \approx \chi \left(\frac{\Gamma\mu}{g}
		\right)^{1/2} \;, 
	\label{eob_chi}
\end{eqnarray}
where
\begin{eqnarray}
	\chi \equiv \frac{E_\ob}{E_{\ph,0}} \;, \hspace{1cm}
		E_{\ph,0} \equiv 3^{-1/4} k_\bo T_0\Gamma \;.
	\label{chi}
\end{eqnarray}

Let us define
\begin{eqnarray}
	R_0\equiv \Gamma c\eta_0 \;. \label{r0}
\end{eqnarray}
Then, $r_\ph/R_0 \approx \eta/\eta_0 = \mu g/\Gamma$. By 
equation (\ref{feo4}) we have
\begin{eqnarray}
	E_\ob F_{E_\ob} &\approx& \frac{10\sigma_\sb T_0^4R_0^2}
		{\pi^4 D^2}\Gamma^2\mu^2 \chi^4 \nonumber\\
		&&\times \int_0^{\Gamma^{-1}}\frac{g^2 \sin\theta\cos\theta 
		d\theta}{\exp\left[(\Gamma\mu/g)^{1/2}\chi\right]-1} \;.
	\label{feo5}
\end{eqnarray}

Define $s\equiv g/\Gamma$ and
\begin{eqnarray}
	\Psi_E \equiv \frac{E_\ob F_{E_\ob}}{F_0} \;, \hspace{1cm}
	F_0\equiv \frac{L_0}{4\pi D^2} \;,\label{psi}
\end{eqnarray}
where $L_0$ is the constant luminosity in the limit $\eta^2/\eta_0^2\ll 1$
given by equation (\ref{lum_th0}) below. By equation (\ref{g_a}), we have 
$\cos\theta \approx \beta_\ph^{-1}\left[1-1/(\Gamma^2 s)\right]$ and
$d\cos\theta \approx \beta_\ph^{-1} ds/(\Gamma^2 s^2)$. When $\theta = 0$,
we have $s = 2$. When $\theta\approx 1/\Gamma$, we have $s\approx 1$.
Hence, equation (\ref{feo5}) can be rewritten as
\begin{eqnarray}
	\Psi_E \approx \frac{45}{7\pi^4} \mu^2 \chi^4 \int_1^2
		\frac{ds}{\exp\left[(\mu/s)^{1/2}\chi\right]-1} \;.
	\label{psi2}
\end{eqnarray}

In equation (\ref{psi2}), $\chi$ and $\mu$ appear in the combination 
$\mu^{1/2}\chi$. Hence, $\Psi_E$ is invariant under the transformation
\begin{eqnarray}
	\mu\rightarrow\mu^\prime \;, \hspace{1cm}
	\chi\rightarrow\chi^\prime = \chi\left(\frac{\mu}{\mu^\prime}
		\right)^{1/2} \;.
	\label{psie_inv}
\end{eqnarray}
That is, the value of $\Psi_E$ at energy $\chi$ and at time $\mu$, is equal 
to the value of $\Psi_E$ at energy $\mu^{1/2}\chi$ and at time $\mu =1$. When 
$\mu=1$, $\Psi_E(\chi)$ peaks at $\chi=4.913$. Hence, at any $\mu$, 
$\Psi_E(\chi)$ peaks at $\chi_\p=4.913\,\mu^{-1/2}$.

Similarly, for the luminosity in equation (\ref{lum_th}), we have
\begin{eqnarray}
	L &\approx& \frac{8\pi}{3}\sigma_\sb T_0^4 R_0^2 \int_0^{\Gamma^{-1}} 
		g^4 \sin\theta\cos\theta d\theta \nonumber\\
		&\approx& L_0\equiv \frac{56\pi}{9}\sigma_\sb T_0^4 R_0^2 
		\Gamma^2 \;,
	\label{lum_th0}
\end{eqnarray}
which is a constant. 

For the photon rate in equation (\ref{pnum_th}), we have
\begin{eqnarray}
	N &\approx& \frac{80\cdot 3^{1/4}\zeta(3)\sigma_\sb}
		{\pi^3 k_\bo}T_0^3 R_0^2 \nonumber\\
		&&\times \int_0^{\Gamma^{-1}} g^3 \left(\frac{\eta}{\eta_0}
		\right)^{1/2} \sin\theta\cos\theta d\theta \nonumber\\
		&=& N_0\mu^{1/2} \;,
	\label{pnum_th0}
\end{eqnarray}
where
\begin{eqnarray}
	N_0 &\equiv& \left(4\sqrt{2}-1\right)\frac{32\cdot 3^{1/4} \zeta(3)
		\sigma_\sb}{\pi^3k_\bo}T_0^3 R_0^2 \Gamma  \nonumber\\
		&=& \left(4\sqrt{2}-1\right)
		\frac{36\zeta(3)}{7\pi^4}\frac{L_0}{E_{\ph,0}} \;.
	\label{pnum0}
\end{eqnarray}

Equations (\ref{lum_th0}), (\ref{psi2}), and (\ref{pnum_th0}) approximate the 
numerical results presented in Figs.~\ref{lightcurve}--\ref{peak_energy} 
very well for $\mu \la 0.1$, with fractional errors $\la 5$ percent.

The integration of $\Psi_E$ over $\mu$ is
\begin{eqnarray}
	\int_0^\mu\Psi_E d\mu = \frac{45}{7\pi^4} \chi^{-2}\int_1^2 ds 
		\int_0^{\mu\chi^2}\frac{w^2 dw}{\exp\sqrt{w/s}-1} \;,
	\label{psi3}
\end{eqnarray}
where $w\equiv \mu \chi^2$.

As $\chi\rightarrow 0$, we have
\begin{eqnarray}
	\int_0^\mu\Psi_E d\mu \approx \frac{12}{7\pi^4}\left(2^{3/2}
		-1\right)\,\mu^{5/2} \chi^3\propto \chi^3 \;.
	\label{chi_0}
\end{eqnarray}
As $\chi\rightarrow\infty$, we have 
\begin{eqnarray}
	\int_0^\mu\Psi_E d\mu \approx \frac{300\pi^2}{49}\,\chi^{-2} \propto 
		\chi^{-2} \;.
	\label{chi_inf}
\end{eqnarray}

\bsp

\label{lastpage}


\begin{thebibliography}{99}

\bibitem[\protect\citeauthoryear{Alpher \& Herman}{1948}]{alp48}
	Alpher R. A., Herman R., 1948, Nat, 162, 774

\bibitem[\protect\citeauthoryear{Band et al.}{1993}]{ban93}
	Band D. et al., 1993, ApJ, 413, 281

\bibitem[\protect\citeauthoryear{Bellm et al.}{2006}]{bel06}
	Bellm E., Bandstra M., Boggs S., Wigger C., Hajdas W., Smith D. M., 
	Hurley K., 2006, GCN 5838

\bibitem[\protect\citeauthoryear{Bennett et al.}{1996}]{ben96}
	Bennett C. L. et al., 1996, ApJ, 464, L1

\bibitem[\protect\citeauthoryear{Bianco et al.}{2001}]{bia01}
	Bianco C. L., Ruffini R., Xue S.-S., 2001, A\&A, 368, 377

\bibitem[\protect\citeauthoryear{Burrows et al.}{2007}]{bur07}
	Burrows D. N. et al., 2007, Phil. Trans. R. Soc. A, 365, 1213

\bibitem[\protect\citeauthoryear{Cenko et al.}{2006}]{cen06}
	Cenko S. B. et al., 2006, ApJ, 652, 490

\bibitem[\protect\citeauthoryear{Daigne \& Mochkovitch}{2002}]{dai02}
	Daigne F., Mochkovitch R., 2002, MNRAS, 336, 1271

\bibitem[\protect\citeauthoryear{Felten \& Rees}{1972}]{fel72}
	Felten J. E., Rees M. J., 1972, A\&A, 17, 226

\bibitem[\protect\citeauthoryear{Fenimore}{2006}]{fen06}
	Fenimore E. et al., 2006, GCN 5831

\bibitem[\protect\citeauthoryear{Friedman \& Bloom}{2005}]{fri05}
	Friedman A. S., Bloom J. S., 2005, ApJ, 627, 1

\bibitem[\protect\citeauthoryear{Gamow}{1948a}]{gam48a}
	Gamow G., 1948a, Phys. Rev., 74, 505

\bibitem[\protect\citeauthoryear{Gamow}{1948b}]{gam48b}
	Gamow G., 1948b, Nat, 162, 680

\bibitem[\protect\citeauthoryear{Giannios}{2006}]{gia06}
	Giannios D., 2006, A\&A, 457, 763

\bibitem[\protect\citeauthoryear{Giannios \& Spruit}{2007}]{gia07}
	Giannios D., Spruit H. C., 2007, A\&A, in press 
	(arXiv:astro-ph/0611385v2)

\bibitem[\protect\citeauthoryear{Golenetskii et al.}{2006}]{gol06}
	Golenetskii S., Aptekar R., Mazets E., Pal'shin V., Frederiks D., 
	Cline T., 2006, GCN 5837

\bibitem[\protect\citeauthoryear{Goodman}{1986}]{goo86}
	Goodman J., 1986, ApJ, 308, L47.

\bibitem[\protect\citeauthoryear{Jackson}{1999}]{jac99}
	Jackson J. D., 1999, Classical Electrodynamics. John Wiley \& Sons,
	New York

\bibitem[\protect\citeauthoryear{Kobayashi et al.}{1999}]{kob99}
	Kobayashi S., Piran T., Sari R., 1999, ApJ, 513, 669

\bibitem[\protect\citeauthoryear{Koshut et al.}{1995}]{kos95}
	Koshut T. M., Kouveliotou C., Paciesas W. S., van Paradijs J., 
	Pendleton G. N., Briggs M. S., Fishman G. J., Meegan C. A., 1995,
	ApJ, 452, 145

\bibitem[\protect\citeauthoryear{Lazzati}{2005}]{laz05}
	Lazzati D., 2005, MNRAS, 357, 722

\bibitem[\protect\citeauthoryear{Lyutikov \& Usov}{2000}]{lyu00}
	Lyutikov M., Usov V. V., 2000, ApJ, 543, L129

\bibitem[\protect\citeauthoryear{MacFadyen \& Woosley}{1999}]{mac99}
	MacFadyen A. I., Woosley S. E., 1999, ApJ, 524, 262

\bibitem[\protect\citeauthoryear{MacFadyen et al.}{2001}]{mac01}
	MacFadyen A. I., Woosley S. E., Heger A., 2001, ApJ, 550, 410

\bibitem[\protect\citeauthoryear{Marcinkowski et al.}{2006}]{mar06}
	Marcinkowski R., Denis M., Bulik T., Goldoni P., Laurent Ph., Rau A.,
	2006, A\&A, 452, 113

\bibitem[\protect\citeauthoryear{McBreen et al.}{2006}]{mcb06}	
	McBreen S., Hanlon L., McGlynn S., McBreen B., Foley S., Preece R., 
	von Kienlin A., Williams O. R., 2006, A\&A, 455, 433

\bibitem[\protect\citeauthoryear{M\'esz\'aros}{2006}]{mes06}
	M\'esz\'aros P., 2006, Rept. Prog. Phys., 69, 2259

\bibitem[\protect\citeauthoryear{M\'esz\'aros \& Rees}{2000}]{mes00}
	M\'esz\'aros P., Rees M. J., 2000, ApJ, 530, 292

\bibitem[\protect\citeauthoryear{M\'esz\'aros et al.}{2002}]{mes02}
	M\'esz\'aros P., Ramirez-Ruiz E., Rees M. J., Zhang B., 2002,
	ApJ, 578, 812

\bibitem[\protect\citeauthoryear{Misner et al.}{1973}]{mis73} 
	Misner C. W., Thorne K. S., Wheeler J. A., 1973, Gravitation.
	Freeman, San Francisco

\bibitem[\protect\citeauthoryear{Murakami et al.}{1991}]{mur91}
	Murakami T., Inoue H., Nishimura J., van Paradijs J., Fenimore E. E.,
	1991, Nat, 350, 592

\bibitem[\protect\citeauthoryear{Murakami et al.}{1992}]{mur92}	
	Murakami T., Ogasaka Y., Yoshida A., Fenimore E. E., 1992, in Paciesas
	W. S., Fishman G. J., eds, AIP Conference Proceedings, Vol. 265, 
	Gamma-Ray Bursts. Am. Inst. Phys., New York, p. 28

\bibitem[\protect\citeauthoryear{Novikov \& Thorne}{1973}]{nov73}
	Novikov I. D., Thorne K. S., 1973, in DeWitt C., DeWitt B. S., eds, 
	Black Holes. Gordon and Breach, New York, p. 343

\bibitem[\protect\citeauthoryear{Paczy\'nski}{1986}]{pac86}
	Paczy\'nski B., 1986, ApJ, 308, L43

\bibitem[\protect\citeauthoryear{Paczy\'nski}{1990}]{pac90}
	Paczy\'nski B., 1990, ApJ, 363, 218

\bibitem[\protect\citeauthoryear{Paczy\'nski \& Xu}{1994}]{pac94}
	Paczy\'nski B., Xu G., 1994, ApJ, 427, 708

\bibitem[\protect\citeauthoryear{Page et al.}{2006}]{pag06}
	Page K. L. et al., 2006, GCN 5823

\bibitem[\protect\citeauthoryear{Page et al.}{2007}]{pag07}
	Page K. L. et al., 2007, ApJ, in press (arXiv:0704.1609v1 [astro-ph])

\bibitem[\protect\citeauthoryear{Peebles}{1993}]{pee93}
	Peebles P. J. E., 1993, Principles of Physical Cosmology. Princeton
	Univ. Press, Princeton

\bibitem[\protect\citeauthoryear{Pe'er et al.}{2006}]{pee06}
	Pe'er A., M\'esz\'aros P., Rees M. J., 2006, ApJ, 642, 995

\bibitem[\protect\citeauthoryear{Pe'er et al.}{2007}]{pee07}
	Pe'er A., Ryde F., Wijers R. A. M. J., M\'esz\'aros P., Rees M. J., 
	2007, ApJ Letters, submitted (arXiv:astro-ph/0703734v1)

\bibitem[\protect\citeauthoryear{Penzias \& Wilson}{1968}]{pen68}
	Penzias A. A., Wilson R. W., 1968, ApJ, 142, 419

\bibitem[\protect\citeauthoryear{Piran}{1999}]{pir99}
	Piran T., 1999, Phys. Rep., 314, 575

\bibitem[\protect\citeauthoryear{Piran}{2004}]{pir04}
	Piran T., 2004, Rev. Mod. Phys., 76, 1143

\bibitem[\protect\citeauthoryear{Ramirez-Ruiz}{2005}]{ram05}
	Ramirez-Ruiz E., 2005, MNRAS, 363, L61

\bibitem[\protect\citeauthoryear{Ramirez-Ruiz et al.}{2002}]{rma02}
	Ramirez-Ruiz E., MacFadyen A. I., Lazzati D., 2002, MNRAS, 331, 197

\bibitem[\protect\citeauthoryear{Rees}{1966}]{ree66}
	Rees M. J., 1966, Nat, 211, 468

\bibitem[\protect\citeauthoryear{Rees \& M\'esz\'aros}{1992}]{ree92}
	Rees M. J., M\'esz\'aros P., 1992, MNRAS, 258, 41P

\bibitem[\protect\citeauthoryear{Rees \& M\'esz\'aros}{1994}]{ree94}
	Rees M. J., M\'esz\'aros P., 1994, ApJ, 430, L93

\bibitem[\protect\citeauthoryear{Rees \& M\'esz\'aros}{2005}]{ree05}
	Rees M. J., M\'esz\'aros P., 2005, ApJ, 628, 847

\bibitem[\protect\citeauthoryear{Rindler}{1977}]{rin77}
	Rindler W., 1977, Essential Relativity: Special, General, and 
	Cosmological. Springer-Verlag, New York	

\bibitem[\protect\citeauthoryear{Romano et al.}{2006}]{rom06}
	Romano P. et al., 2006, A\&A, 456, 917

\bibitem[\protect\citeauthoryear{Ruffini et al.}{2001}]{ruf01}
	Ruffini R., Bianco C. L., Fraschetti F., Xue S.-S., Chardonnet P.,
	2001, ApJ, 555, L113

\bibitem[\protect\citeauthoryear{Ruffini et al.}{2002}]{ruf02}
	Ruffini R., Bianco C. L., Chardonnet P., Fraschetti F., Xue S.-S.,
	2002, ApJ, 581, L19

\bibitem[\protect\citeauthoryear{Ruffini et al.}{2005}]{ruf05}
	Ruffini R., Bernardini M. G., Bianco C. L., Chardonnet P., 
	Fraschetti F., Gurzadyan V., Vitagliano L., Xue S.-S., 2005,
	in Novello M., Bergliaffa S. E. P., eds, AIP Conference Proceedings,
	Vol. 782, Cosmology and Gravitation. Am. Inst. Phys., New York, p. 42

\bibitem[\protect\citeauthoryear{Ryde}{2004}]{ryd04}
	Ryde F., 2004, ApJ, 614, 827

\bibitem[\protect\citeauthoryear{Ryde}{2005}]{ryd05}
	Ryde F., 2005, ApJ, 625, L95
\bibitem[\protect\citeauthoryear{Sari, Piran \& Halpern}{1999}]{sar99}
	Sari R., Piran T., Halpern J. P., 1999, ApJ, 519, L17

\bibitem[\protect\citeauthoryear{Sazonov et al.}{1998}]{saz98}
	Sazonov S. Y., Sunyaev R. A., Terekhov O. V., Lund N., Brandt S., 
	Castro-Tirado A. J., 1998, A\&AS, 129, 1

\bibitem[\protect\citeauthoryear{Shakura}{1972}]{sha72}
	Shakura N. I., 1972, Sov. Astron., 16, 532

\bibitem[\protect\citeauthoryear{Shakura \& Sunyaev}{1973}]{sha73}
	Shakura N. I., Sunyaev R. A., 1973, A\&A, 24, 337

\bibitem[\protect\citeauthoryear{Sunyaev \& Shakura}{1974}]{sun74}
	Sunyaev R. A., Shakura N. I., 1974, Sov. Astron., 18, 60

\bibitem[\protect\citeauthoryear{Shemi \& Piran}{1990}]{she90}
	Shemi A., Piran T., 1990, ApJ, 365, L55

\bibitem[\protect\citeauthoryear{Smoot et al.}{1992}]{smo92}
	Smoot G. F. et al., 1992, ApJ, 396, L1

\bibitem[\protect\citeauthoryear{Thompson}{1994}]{tho94}
	Thompson C., 1994, MNRAS, 270, 480

\bibitem[\protect\citeauthoryear{Thompson et al.}{2007}]{tho07}
	Thompson C., M\'esz\'aros P., Rees M. J., 2007, ApJ, submitted
	(arXiv:astro-ph/0608282v2)

\bibitem[\protect\citeauthoryear{Vestrand et al.}{2005}]{ves05}
	Vestrand W. T. et al., 2005, Nat, 435, 178

\bibitem[\protect\citeauthoryear{Waxman \& M\'esz\'aros}{2003}]{wax03}
	Waxman E., M\'esz\'aros P., 2003, ApJ, 584, 390

\bibitem[\protect\citeauthoryear{Woosley \& Bloom}{2006}]{woo06}
	Woosley S. E., Bloom J. S., 2006, ARA\&A, 44, 507

\bibitem[\protect\citeauthoryear{Zhang \& M\'esz\'aros}{2004}]{zha04}
	Zhang B., M\'esz\'aros P., 2004,  Int. J. Mod. Phys. A, 19, 2385

\end{thebibliography}
\end{document}